\documentclass[twocolumn,showpacs,amsmath,amssymb]{revtex4}

\usepackage{amssymb}
\usepackage{amsmath}
\usepackage[english]{babel}
\usepackage{graphicx}
\usepackage{times}

\newcommand{\be}{\begin{equation}}  
\newcommand{\ee}{\end{equation}}

\newcommand{\bea}{\begin{eqnarray}}           
\newcommand{\eea}{\end{eqnarray}} 
\newcommand{\ba}{\begin{align}}
\newcommand{\ea}{\end{align}}
\def\de{\partial}
\def\be{\begin{equation}}
\def\ee{\end{equation}}
\def\beq{\begin{eqnarray}}
\def\eeq{\end{eqnarray}}
\newcommand{\beqn}{\begin{eqnarray*}}
\newcommand{\eeqn}{\end{eqnarray*}}
\def\lm{{\ell m}}

\def\l{{\ell }}

\def\IL{\relax{\rm I\kern-.18em L}}

\begin{document}

\title{On the accretion-induced QNM excitation of a Schwarzschild black hole}

\author{Alessandro Nagar$^{1,3}$, Olindo Zanotti$^{2,3}$, Jos\'e A. Font$^{3}$ and
  Luciano Rezzolla$^{4,5,6}$ } 
      
\affiliation{ $^1$Dipartimento di Fisica and INFN, Politecnico di Torino,
                   Torino, Italy}

\affiliation{ $^2$Dipartimento di Astronomia e Scienza dello Spazio
                   Universit\`a di Firenze, Firenze, Italy }

\affiliation{$^3$Departamento de Astronom\'ia y Astrof\'isica,
                    Universidad de Valencia,
                    Valencia, Spain}   

\affiliation{$^4$Max-Planck-Institut f\"ur Gravitationsphysik,
  Albert-Einstein-Institut, Potsdam-Golm, Germany}

\affiliation{$^5$SISSA, International School for Advanced Studies and
  INFN, Trieste, Italy }

\affiliation{$^6$Department of Physics, Louisiana State University,
  Baton Rouge, USA }

\date{\today}

\begin{abstract}
  By combining the numerical solution of the nonlinear hydrodynamics
  equations with the solution of the linear inhomogeneous
  Zerilli-Moncrief and Regge-Wheeler equations we investigate the
  properties of the gravitational radiation emitted during the
  axisymmetric accretion of matter onto a Schwarzschild black
  hole. The matter models considered include quadrupolar dust shells
  and thick accretion disks, permitting us to simulate situations
  which may be encountered at the end stages of stellar gravitational
  collapse or binary neutron star merger. We focus on the interference
  pattern appearing in the energy spectra of the emitted gravitational
  waves and on the amount of excitation of the quasi-normal modes of
  the accreting black hole. We show that, quite generically in the
  presence of accretion, the black hole ringdown is not a simple
  superposition of quasi-normal modes, although the fundamental mode
  is usually present and often dominates the gravitational-wave
  signal. We interpret this as due to backscattering of waves off the
  non-exponentially decaying part of the black hole potential and to
  the finite spatial extension of the accreting matter. Our results
  suggest that the black hole QNM contributions to the full
  gravitational wave signal should be extremely small and possibly not
  detectable in generic astrophysical scenarios involving the
  accretion of extended distributions of matter.
\end{abstract}

\pacs{
%04.25.Dm,  % numerical relativity
04.30.Db,   % gravitational-wave generation and sources
04.40.Dg,   % Relativistic stars: structure, stability, and oscillations
%04.70.Bw,  % classical black holes
%95.30.Sf,  % relativity and gravitation
95.30.Lz,   % Hydrodynamics
%97.60.Jd%, % Neutron stars
%97.60.Lf   % black holes (astrophysics)
98.62.Mw    % Infall, accretion, and accretion discs
}

\maketitle

%%%%%%%%%%%%%%%%%%%%%%
\section{Introduction}
%%%%%%%%%%%%%%%%%%%%%%

It is well known within the framework of black-hole perturbation
theory~\cite{chandra83,fn_98,KokkotasSchmidt} that quasi-normal modes
(QNMs) (i.e., exponentially damped harmonic oscillations)
dominate the gravitational-wave response of a non-spherically distorted
black hole if the corresponding frequencies are part of the Fourier
spectrum of the external source that moved the hole away from its
equilibrium state. This perturbation is then radiated away in form of
gravitational radiation until the black hole returns to its
unperturbed, quiescent state. In practice, the QNMs excitation is
triggered if the frequency of the perturbing agent (e.g. an external
matter source moving close to the black hole) is sufficiently close to
the fundamental frequencies of the black hole, which then acts as an
excited oscillator. As a result, the QNMs and in particular the
fundamental mode (which is the one at the highest frequency and with
the smallest damping time) represent the main feature of the
gravitational waves emission only for a sufficiently compact
perturbation; i.e., for sources whose characteristic scale is
comparable with the width of the peak of the potential.

These results are well-known since the early studies of
Press~\cite{press72} (see also Vishveshwara~\cite{vish70}), who considered
the scattering of Gaussian gravitational-wave packets off a
Schwarzschild black hole and noticed that the excitation of the QNMs
of the black hole is more efficient for very narrow
packets. Approximate relations to compute the efficiency of excitation
of the various QNMs were introduced by Leaver~\cite{leaver86} and by
Andersson~\cite{andersson95a} for Gaussian pulses initial data with
variable width. Since the QNMs spectrum is entirely determined by the
black hole properties (i.e., mass, spin and charge), it is
expected that the detection of the QNMs ringdown would provide a
unique opportunity to unveil the physical properties of a black hole.
For this reason, the excitation of black hole QNMs have been studied
in various astrophysical scenarios, such as the gravitational collapse
of a (rotating) neutron star to a black hole or the collision of two
black holes. The presence of the QNMs in these situations has been
confirmed either through the use of perturbation theory with various
degrees of sophistication~\cite{CPMI, seidel91, lousto97a}, or through
fully relativistic numerical
simulations~\cite{baiotti05a,baiotti06,pretorius05a,campanelli05a,baker05a}.

However, under more general and realistic conditions, such as the
excitation of the QNMs by accretion of matter, and that may be
encountered in gravitational stellar collapse or binary neutron star
mergers, the gravitational-wave response of a black hole can be more
complex. For instance, the simulations performed
by~\cite{PapadopoulosFont,nagar04,nagar05a} showed that the
gravitational-wave signal is not simply given by the superposition of
exponentially-damped sinusoids at the QNMs frequencies and that, in
some cases, the QNM ringing is only weakly excited and analysis in the
frequency domain are needed~\cite{ferrari06}. On the other hand,
``backscattering'' effects related to the slowly-decaying features of
the scattering potential and interference effects turn out to play a
crucial role for the correct interpretation of the results; a detailed
discussion of these effects can be found in Ref.~\cite{nagar05a}.

In the case of a Schwarzschild black hole of mass $M$, we recall that 
the appearance of QNMs ringing is related to the peak of the curvature
potential (also referred to as the Zerilli or Regge-Wheeler
potential), that is located at $r\simeq 3\,M$; the
backscattering, on the other hand, is related to the fact that the
curvature potential decays as $r^{-2}$ for $r\rightarrow\infty$. 
This behavior is responsible for the late time power-law tail 
$t^{-(2\ell +3)}$ of the gravitational-wave signal, where $\ell$ 
refers to the radiation multipole.

Therefore, while early-time ringing is the result of a superposition
of exponentially damped sinusoids, and is dominated by the
fundamental quasi-normal mode, at later times the ringing dies
out and the signal is dominated by tail effects. However, in the
transition from the ``ringdown'' to the ``tail'' phase, additional
oscillations appear that cannot be attributed to any of the two
regimes and that also seem to depend on the choice of the initial data
(see~\cite{KokkotasSchmidt,fn_98}). As we shall show in this paper for
a broad sample of initial data, there could be intermediate regimes
where the ``ringdown'' and the ``tail'' terms of the potential produce
competing effects, so that the QNMs ringing and the backscattering
effects can overlap, generating complex waveforms. These effects were
first noticed in Refs.~\cite{PapadopoulosFont,nagar04}, although not
discussed in detail there.

In a recent paper~\cite{nagar04}, hereafter Paper I, a general
analysis of the gravitational radiation emitted as a result of
anisotropic accretion of matter shells onto nonrotating black holes
and neutron stars was presented.  That investigation made use of a
procedure that combines the solution of the linearized Einstein
equations for the metric perturbations with fully nonlinear
hydrodynamics simulations. Although the study of black hole
perturbations produced by infalling matter has a long history and rich
literature~\cite{sw_82,psw_85,sotani05a}, the approach outlined in
Paper I proved to be useful for a number of reasons: {\em i)} it
provided additional information on the black-hole's response to the
dynamics of point-like particles in the vicinity of black holes; {\em
  ii)} it helped understanding the basic black-hole's response to
extended matter perturbations; {\em iii)} it represented an effective
way of studying black-hole physics in a linear regime without having
to resort to full-scale numerical relativity simulations.

One of the main results of Paper I was that, in the idealized
accretion processes considered, most of the energy is released at
frequencies lower than that of the fundamental quasi-normal mode (QNM)
of the black hole, the spectrum consisting of a complex pattern,
mostly produced during the accretion process rather than in the
ringdown phase. More precisely, the gravitational-wave emission was
found to be dominated by a collection of interference ``fringes'' at
frequencies of about a few hundred Hz, rather than by a single
monochromatic peak at the (higher) frequency of the fundamental mode
of the black hole. Moreover, the width of these fringes was found to
decrease rapidly with the initial position of the matter source. These
results, which were already observed in other
works~\cite{lousto97a,martel01} in the case of a point-like particle
falling onto a Schwarzschild black hole, also showed that the
appearance of interference fringes in the energy spectra is much
larger when the accreting matter is a shell of finite size and that
the efficiency in gravitational-wave emission is much reduced,
becoming almost two orders of magnitude smaller than in the case of
point-like particles. An important feature of the calculations carried
out in Paper I was the minimization of the initial gravitational-wave
content; this turned out to be crucial to illustrate that the
interference pattern was mainly due to the finite radial extension of
the accreting source.

The aim of the present paper is twofold. Firstly, we intend to
complete the discussion started in Paper I on accreting quadrupolar
shells onto a Schwarzschild black hole by extending the parameter
space of the initial models and by analyzing their impact on the
gravitational-wave emission. In particular we study how the energy
emitted in gravitational waves, and the corresponding spectra, depend
on the compactness of the shells as well as on their initial
locations. In doing so we show that, for a finite-size source, the
ringing of the black hole is much more complex than a simple
superposition of QNMs and that the energy spectra (and in particular
the interference fringes) are dependent on the choice of the initial
data.

Secondly, we improve the astrophysical relevance of our study by
analyzing the gravitational radiation produced by thick accretion
disks~\cite{abramovicz78,font02a} which accrete onto the black hole on
dynamical timescales. We recall, in fact, that quasi-periodic
oscillations of thick accretion disks (or tori) orbiting around
Schwarzschild or Kerr black holes have been recently addressed as
promising sources of gravitational
waves~\cite{nagar05a,zanotti03,zanotti05,rezzollaetal_03a,rezzollaetal_03b} 
in the kHz range.

The paper is organized as follows: in Sec.~\ref{analytic} we review
the theory of odd- and even-parity nonspherical perturbations of
Schwarzschild spacetime, writing the inhomogeneous Zerilli-Moncrief
and Regge-Wheeler equations in a form suitable for time-domain
calculations. Sec.~\ref{num_app} briefly describes the numerical
approach adopted for the simulations, while Sec.~\ref{results} is
devoted to the discussion of the results. Finally,
Sec.~\ref{conclusions} provides a summary of the most important
results and presents our conclusions. Unless otherwise specified, we
choose geometrized units ($c=G=1$), and the black hole mass $M$ is the
unit of length.

%%%%%%%%%%%%%%%%%%%%%%%%%%%%%%%%%%%%%%%%%%%%%%%%%%%%%%%%%%%%%%%
\section{Gauge-invariant perturbations of the Schwarzschild metric}
\label{analytic}
%%%%%%%%%%%%%%%%%%%%%%%%%%%%%%%%%%%%%%%%%%%%%%%%%%%%%%%%%%%%%%%
%
%-------------------------------------------------------------
\subsection{Odd and even-parity master equations}
%-------------------------------------------------------------

The theory of gauge-invariant nonspherical metric perturbations of a
Schwarzschild spacetime has a long history which has been recently
reviewed in Refs.~\cite{mp_05,nagar04b}. We here simply recall that in
this approach the spacetime metric $g_{\mu\nu}$ is described by the
background Schwarzschild metric $g{\!\!\!\!\;
  \raisebox{-0.1ex}{$^{^0}$}}_{\mu\nu}$ plus a nonspherical
perturbation $h_{\mu\nu}$. As a result of the spherical symmetry,
$h_{\mu\nu}$ can be expanded in three odd-parity and seven even-parity
multipoles
\begin{equation}
g_{\mu\nu}=g{\!\!\!\!\; \raisebox{-0.1ex}{$^{^0}$}}_{\mu\nu}
          +\sum_{\l=0}^{\infty}\sum_{m=-\l}^{\l}
          \left(h_{\mu\nu}^{\lm}\right)^{(\mathrm{o})}
          +\left(h_{\mu\nu}^{\lm}\right)^{(\mathrm{e})}\;,
\end{equation}
with the odd multipoles transforming as $(-1)^{\l+1}$ and the even
ones as $(-1)^{\l}$ under a parity transformation
$(\theta,\varphi)\rightarrow(\pi-\theta,\pi+\varphi)$.  The presence
of matter around the black hole is accounted through a ``source'' term
in the linearized Einstein's equations, represented by a stress energy
tensor $t_{\mu\nu}$ which can also be expanded in multipoles.  From
the multipoles of the perturbed metric, it is possible to build odd,
$\Psi^{(\rm o)}_{\lm}$, and even-parity, $\Psi^{(\rm e)}_{\lm}$, gauge
invariant quantities which are solution of two equations, the
Regge-Wheeler~\cite{ReggeWheeler} and
Zerilli-Moncrief~\cite{Zerilli,Moncrief} equations, respectively and
whose expression in Schwarzschild coordinates is given by
\begin{equation}
\label{R-W}
 \de_{t}^2 \Psi^{(\rm o/e)}_{\lm}-\de_{r_*}^2\Psi^{(\rm
o/e)}_{\lm}+V_{\l}^{(\rm o/e)}\Psi^{({\rm o/e})}_{\lm} = S^{(\rm o/e)}_{\lm} \ .
\end{equation}
Here the upper indices refer to the odd and even case respectively and
$r_*=r+2M\log[r/(2M)-1)] $ is the ``tortoise'' coordinate 
(see Refs.~\cite{mp_05,st01} for the most general covariant form of these equations). 
Explicit expressions for the scattering potentials $V_{\l}^{({\rm o})}$ and
$V_{\l}^{({\rm e})}$ are well known and can be found, for example, in
Ref.~\cite{nagar04b}. The source terms in Schwarzschild coordinates
can also be found in~\cite{nagar04,nagar04b}, while the corresponding
expressions valid for any coordinate slicing of Schwarzschild
spacetime are reported in detail in Ref.~\cite{mp_05}.
Once $\Psi^{(\rm e)}_{\ell m}$ and $\Psi^{(\rm o)}_{\ell m}$ are
known, the gravitational-wave amplitude can be computed as
\begin{equation}
\label{eq:GW_amplitude}
h_+-{\rm i}h_{\times}=\frac{1}{r}\sum_{\l\geq 2,m}\sqrt{\frac{(\l+2)!}{(\l-2)!}}
   \left(\Psi^{(\rm e)}_{\lm}
  +{\rm i}\Psi^{(\rm o)}_{\lm}\right)
  \;_{-2}Y^{\lm} \ ,
\end{equation}
where $\;_{-2}Y^{\lm}\equiv\,_{-2}Y^{\lm}(\theta,\varphi)$ are the $2$
spin-weighted spherical harmonics~\cite{goldberg67} and the emitted
power in gravitational waves is simply given by
\begin{equation}
\label{eq:dEdt}
\frac{dE}{dt}=\frac{1}{16\pi}\sum_{\l\geq 2,m}\frac{(\l+2)!}{(\l-2)!}
	\left(\left|\dot{\Psi}^{(\rm o)}_{\lm}\right|^2+
        \left|\dot{\Psi}^{(\rm e)}_{\lm}\right|^2\right)\;,
\end{equation}
where the overdot refers to a derivative with respect to the
Schwarzschild coordinate time.

%-------------------------------------
\subsection{Initial data}
\label{sbsc:id}
%-------------------------------------

Specifying suitable initial data for nonspherical metric perturbations
taking place in astrophysical events is not trivial.  The standard
approach, in the case of even-parity perturbations, exploits the fact
that $\Psi^{(\rm e)}_{\ell m}$ can be written in terms of two 
gauge-invariant multipoles, namely the perturbed conformal factor $k_{\ell m}$ 
and the gravitational-wave degree of freedom $\chi_{\ell m}$ as introduced, 
for instance, in Eq.~(8) of Paper I and whose relationship with the 
Zerilli-Moncrief equation is given by Eq.~(18) of the same 
reference~\cite{nagar04b,GS79}
As a result, the perturbed Hamiltonian equation and the equation for the 
momentum constraint, i.e., Eqs.~(45) and (46) of~\cite{nagar04b}, 
provide a system of coupled ordinary differential equations for the
unknowns ($k_{\ell m}$, $\chi_{\ell m}$, $\de_tk_{\ell m}$, and 
$\de_t\chi_{\ell m}$).  The indetermination inherent in the solution 
of this system can be overcome, for example, 
by assuming $\chi_{\ell m}=\beta k_{\ell m}$~\cite{martel01}, 
where $\beta\geq 0$, so that the Hamiltonian constraint simply 
reads
\begin{align}
\label{be1}
&\partial^2_{r_*} k_{\ell m}+\left[\dfrac{2}{r}-\dfrac{5M}{r^2}-\beta\left(\dfrac{1}{r}-
\dfrac{2M}{r^2}\right)\right]\partial_{r_*} k_{\ell m}\nonumber\\
&-\left(1-\dfrac{2M}{r}\right)\left[\dfrac{\Lambda}{r^2}+
\beta\dfrac{\Lambda+2}{2r^2}\right]k_{\ell m}=-8\pi T^{\lm}_{00} \ \  ,
\end{align}
where $\Lambda=\ell(\ell+1)$ and $T^{\lm}_{00}$ are the multipoles
of the energy density $t_{00}$ of the matter source according to
the multipolar decomposition given by Eq.~(13) of Paper I 
(see also Ref.~\cite{GS79}). The case $\beta=0$ corresponds to 
conformally flat initial data, since in this case the perturbed 
metric in isotropic coordinates is the Schwarzschild one modulo a 
conformal transformation. 

Depending on the problem under investigation, we can either choose
initial conditions that are time-symmetric (i.e., for which
$\de_t k_{\ell m}=\de_t\chi_{\ell m}=0$) or not. In the first case, 
if the Hamiltonian constraint is satisfied, the momentum constraint 
is automatically satisfied too. In the second case, on the other hand, 
the momentum constraint must also be solved for $\de_tk_{\ell m}$ 
if we provide a suitable ansatz for $\de_r\chi_{\ell m}$.

Whether it is possible to use time-symmetric initial data depends
essentially on the astrophysical scenario under investigation and
thus on the form of the stress-energy tensor of the matter, whose
multipoles appear as sources in the Hamiltonian and momentum
constraint equations. Clearly, time-symmetry is satisfied for matter
configurations in equilibrium, or initially at rest and subsequently
falling radially; in this case only the numerical solution of the
Hamiltonian constraint is needed. 

This is the case, for example, of dust shells initially at rest at
a finite radius. Because this configuration is intrinsically non
time symmetric, it will invevitably produce spurious radiation. 
A way to minimize the latter is to freeze the sources of the 
perturbation equations up until the radiation has left the grid.
This is the approach we will follow in most of our simulations.

%%%%%%%%%%%%%%%%%%%%%%%%%%%%%%%%%%%%%%%
\section{Numerical framework}
\label{num_app}
%%%%%%%%%%%%%%%%%%%%%%%%%%%%%%%%%%%%%%%

The numerical approach adopted in our simulations is the same as the
one described in Paper I. We only recall here that our hybrid approach
implies the solution of both the nonlinear,  relativistic hydrodynamics
equations on a fixed Schwarzschild background, in the two spatial 
dimensions $r$ and $\theta$ (since we restrict ourselves to 
axisymmetric flows), and of the
Regge-Wheeler and Zerilli-Moncrief, in the radial direction $r_*$. 
Such ``test-fluid approximation'' is valid as long as the mass of 
the accretion flow is much smaller than the mass of the black hole;
in our simulations this mass ratio is $\lesssim 10\%$

The hydrodynamics equations, once cast in conservation form, are solved
using Godunov-type methods based on Riemann solvers (see
e.g.~\cite{font_lrr} and references therein), while Eqs.~(\ref{R-W})
can be written as a first-order hyperbolic system over a
one-dimensional (1D) grid expressed in terms of the radial coordinate
$r_*$ and solved with some standard finite-difference method such as
Lax-Wendroff (see Paper I for details).

The computational domain used for the hydrodynamical evolution
consists of $N_r\times N_{\theta}$ grid-points, geometrically
distributed along $r$ [that is, with grispacing 
$ \Delta  r_{j+1}=\alpha \Delta r_j$ 
with $\alpha\gtrsim 1$] and uniformly distributed along $\theta$. 
When we evolve quadrupolar shells of dust plunging from large distances 
we set $N_r = 2000$ and $N_{\theta} = 20$, while for the evolution of
fluid accretion disks we use $N_r = 300$ and $N_{\theta} = 150$ to
reach the desired truncation error. 
The hydrodynamical grid extends from a point slightly outside  
the event horizon up to a finite radius,
that depends on the initial data, much smaller than
the extraction radius.
At the inner boundary 
we impose inflowing boundary conditions, while at the outer one we fix 
the conditions of a tenuous stationary spherical atmosphere.

The radial grid used for the time
evolution of the Regge-Wheeler and Zerilli-Moncrief equations, on the
other hand, is much more extended in radius than the hydrodynamical
one and overlaps with the latter. To avoid spurious boundary effects,
the 1D grid typically extends from $-2000\,M$ to $5000\,M$ and is covered
by $\sim 10^5$ cells. At the inner and outer boundary of this grid we 
use standard Sommerfeld outgoing conditions. We stress that, due to the 
long-range action of the potential, these conditions are not completely 
non-reflecting, which is the reason why we still need large grids 
to properly capture the late time power-law decay of the waveforms 
(see below). We recall in passing that exactly non-reflecting
boundary condition for the Zerilli-Moncrief and Regge-Wheeler 
equations have been recently proposed in~\cite{lau05a}.

While the gravitational waveforms are extracted using both the
Regge-Wheeler and Zerilli-Moncrief equations, we also use as a
comparison the Newtonian quadrupole formula to compute the amount of
gravitational waves emitted. We use the modified definition of the
quadrupole moment proposed in Ref.~\cite{nagar05a} and which reads
\begin{equation}
\label{sqf1}
I=2\pi\int_{-1}^{1}dz\int_0^\infty(hDW-p)
	\left(\frac{3}{2}z^2-\frac{1}{2}\right)r^4 dr \ ,
\end{equation}
where $D=\rho W$ is the conserved rest-mass density, $W$ is the
relativistic Lorentz factor, $h$ is the relativistic enthalpy,
$p$ is the pressure, and $z=\cos\theta$. Expression~\eqref{sqf1}
represents a slight modification of the Newtonian standard quadrupole
formula (SQF$_1$) and it was shown to provide the best agreement with the
gauge-invariant waveforms when compared with alternative
definitions~\cite{nagar05a}.

%%%%%%%%%%%%%%%%%%%%%%%%%%%%%%%%%%%%%%%%%%%%%%%%%%%%%%
\section{Results}
\label{results}
%%%%%%%%%%%%%%%%%%%%%%%%%%%%%%%%%%%%%%%%%%%%%%%%%%%%%%
%---------------------------------------------------------------
\subsection{Effects of the gravitational potential}
\label{sec_PT}
%---------------------------------------------------------------

As mentioned in the Introduction, for a certain set of initial
conditions the ringdown phase of the gravitational-wave signal from a
perturbed black hole shows the presence of the first QNMs of the black
hole (typically, the fundamental mode and the first overtone) together
with ``backscattering'' or ``tail''effects. While the less damped
modes are related to the interaction of the perturbations with the
peak of the black hole Regge-Wheeler potential, the tail-effect cannot 
be associated to QNMs and are instead due to the long-range features 
of the potential and in particular to its slow decay with radius.

%--------- FIG.2: comparison between waveforms: PT versus RW potential --------------
\begin{figure*}[t]
\begin{center}
\vspace{-0.5 cm}
\includegraphics[width=8.25 cm]{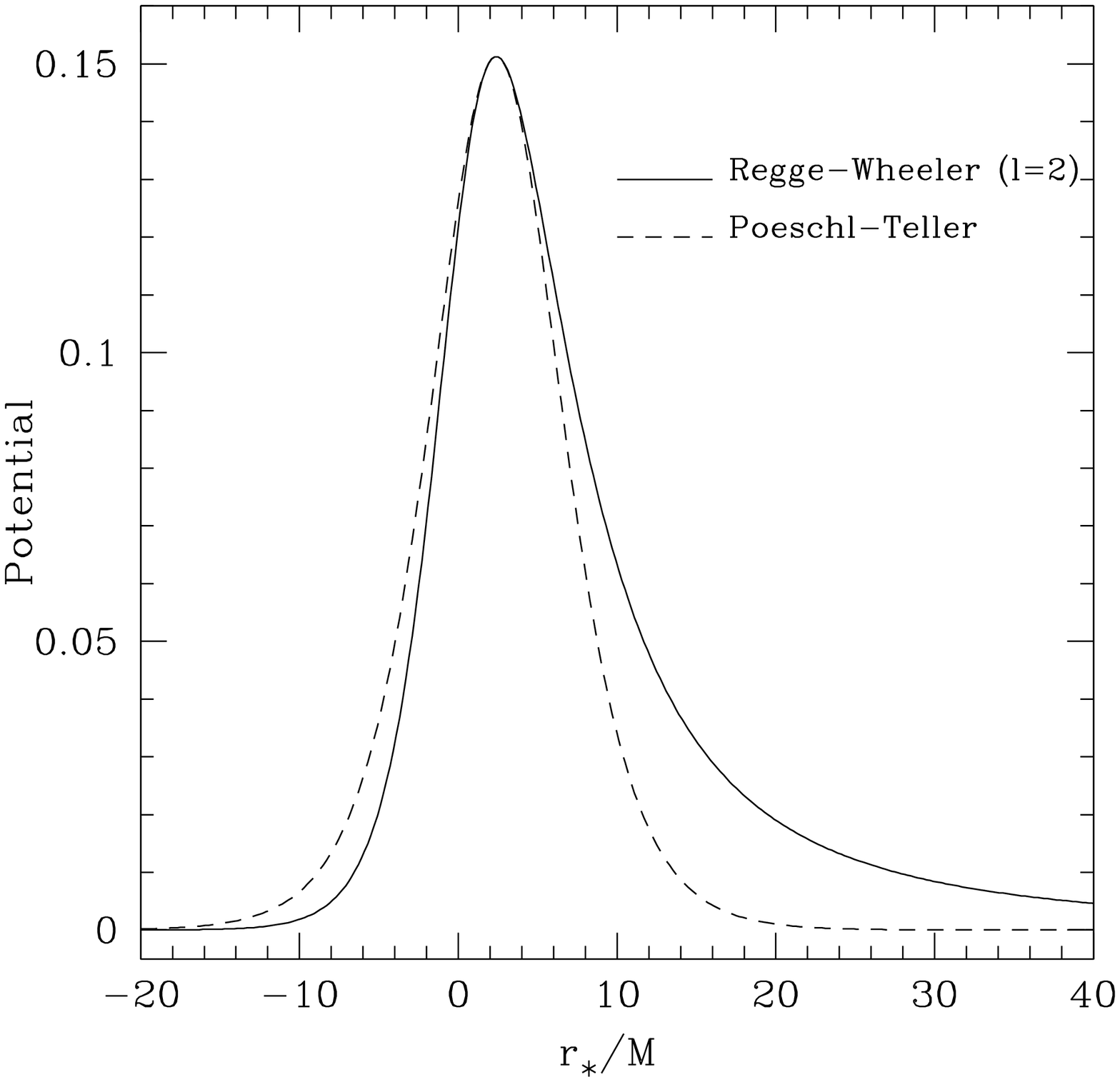}\hspace{0.2 cm}
\hspace{0.5 cm}
\includegraphics[width=8.25 cm]{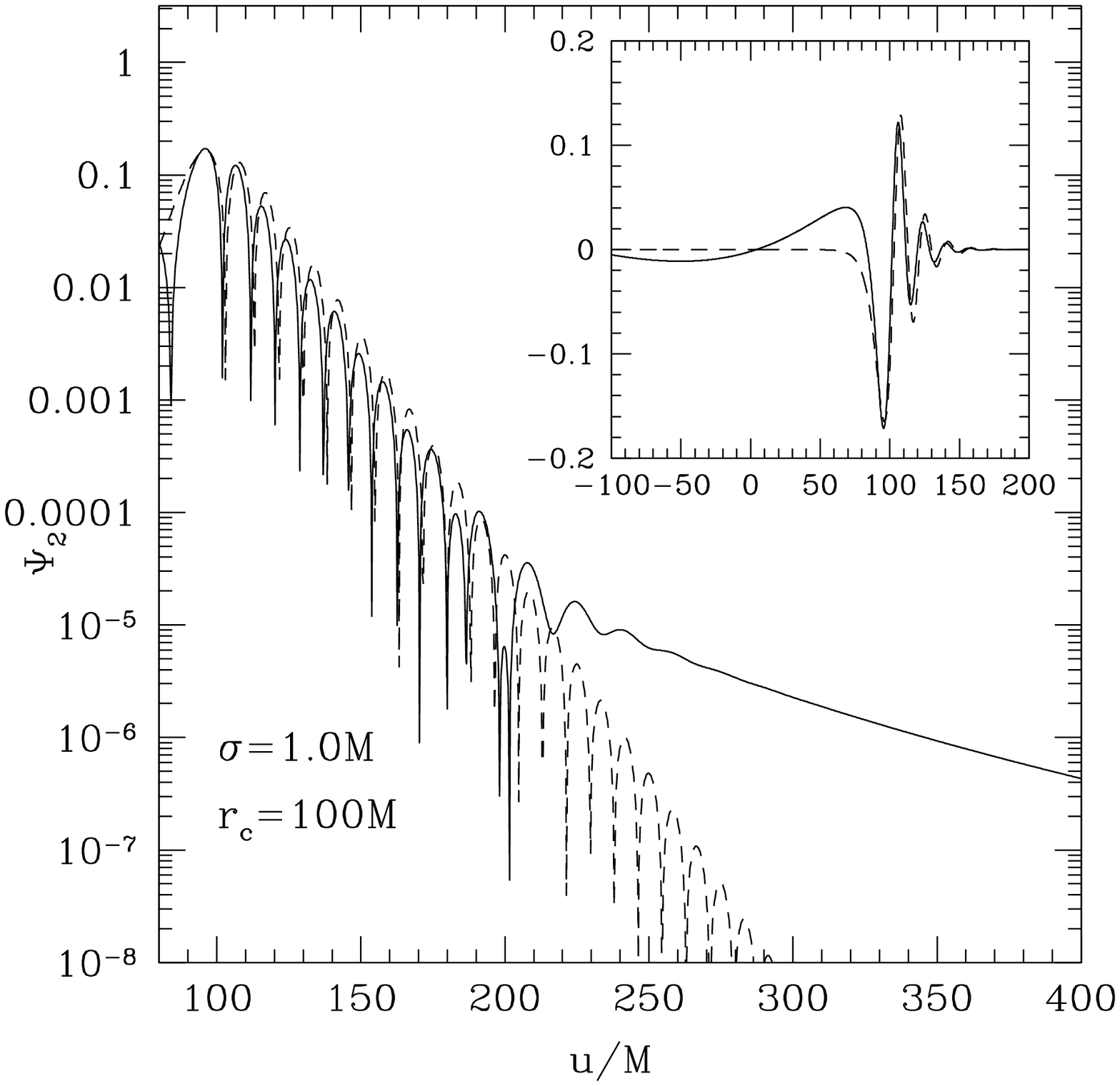} \\
\vspace{-0.5 cm}
\includegraphics[width=8.25 cm]{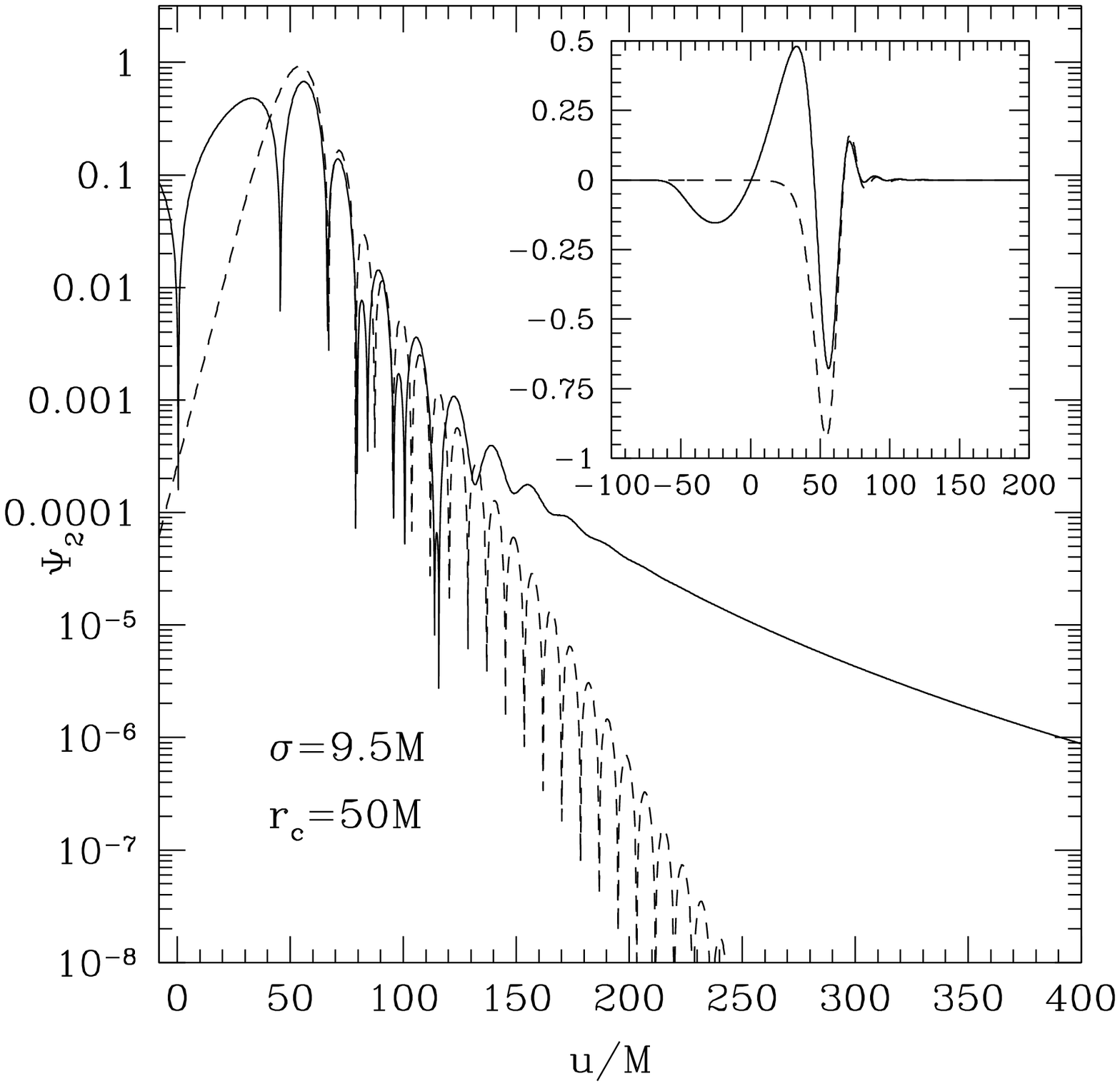}\hspace{0.2 cm}
\hspace{0.5 cm}
\includegraphics[width=8.25 cm]{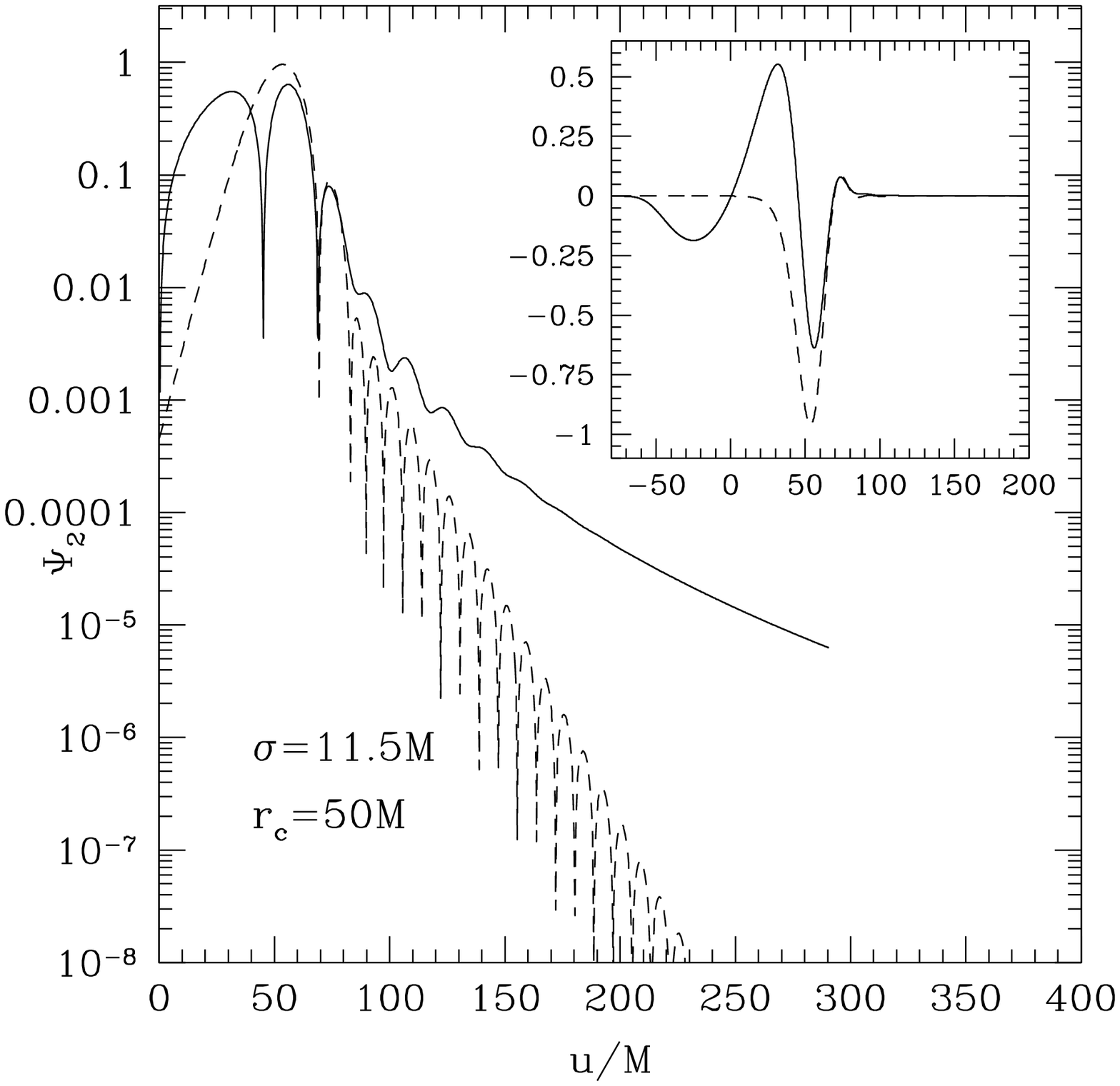}
\caption{\label{fig_2}Comparison between the outcome of a
  scattering problem over the Regge-Wheeler (solid lines) and the
  P\"oschl-Teller (dashed lines) potential. The waveforms are
  extracted at $r_{\rm obs}=200\,M$ and refer to a choice of the
  parameters given by (from top to bottom and from left to right):
  $\sigma=\,M$ and $r_c=100\,M$, $\sigma=9.5\,M$ and $\sigma=11.5\,M$ with
  $r_c=50\,M$.  The wider the pulse, the larger the differences
  between the two solutions. The top left panel shows a comparison
  between the profiles of the Regge-Wheeler and the P\"oschl-Teller 
  potentials.}
\end{center}
\vspace{-0.5 cm}
\end{figure*}
%-------------------------------------------------------------------------------------------------

In order to single out the contribution of the peak of the curvature
potential on the generation of the resulting gravitational-wave
spectrum, while suppressing the effect of the tail term, we solve a
simplified version of Eq.~(\ref{R-W}) in which the Regge-Wheeler
potential is replaced by the so-called P\"oschl--Teller
potential~\cite{pt33a,beyer99a}, which is exponentially decaying for
$r_*\rightarrow\pm\infty$. We therefore evolve in time an equation of
the type
\begin{equation}
\label{wave_equation}
\de_t^2 \Psi_{\ell}  -\de_{r_*}^2\Psi_{\ell} + U_{\ell}\Psi_{\ell} = 0 \ ,
\end{equation}
where $U_{\ell}$ is given by
\begin{equation}
\label{eq:PT}
U_{\ell} \equiv \dfrac{V^{\rm max}_{\ell}}
	{\cosh^2\left[\alpha(r_*-r_*^{\rm max})\right]} \ 
\end{equation}
and $r^{\rm max}_*$ is the position of the peak of the Regge-Wheeler
potential $V^{\rm max}_{\ell}\equiv V^{(\rm o)}_{\ell}(r^{\rm max}_{*})$ 
and $\alpha$ is determined through the  matching of the second derivatives 
at $r^{\rm max}_*$; i.e.,
$d^2U_{\ell}/dr_*^2=d^2V^{(\rm o)}_{\ell}/dr_*^2$ at $r_*=r_*^{\rm max}$. 
In the following, for simplicity we will consider only the $\ell=2$ 
odd-parity perturbations.

%---------- FIG 3: WAVEFORM AND QNMS FIT -----------
\begin{figure*}[t]
\begin{center}
\vspace{-0.5 cm} 
\includegraphics[width=8.25 cm]{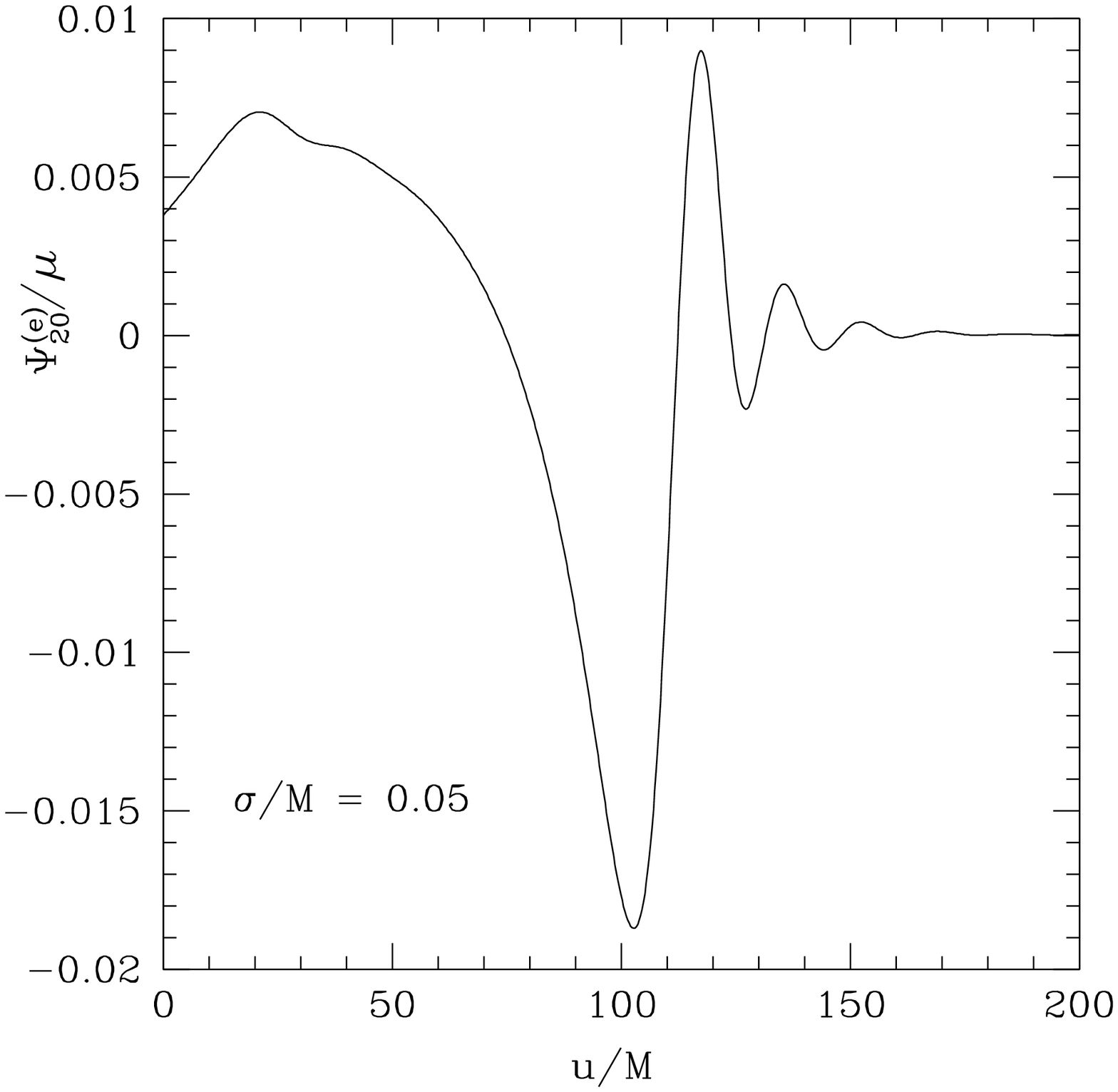}
\hspace{0.5 cm}
\includegraphics[width=8.25 cm]{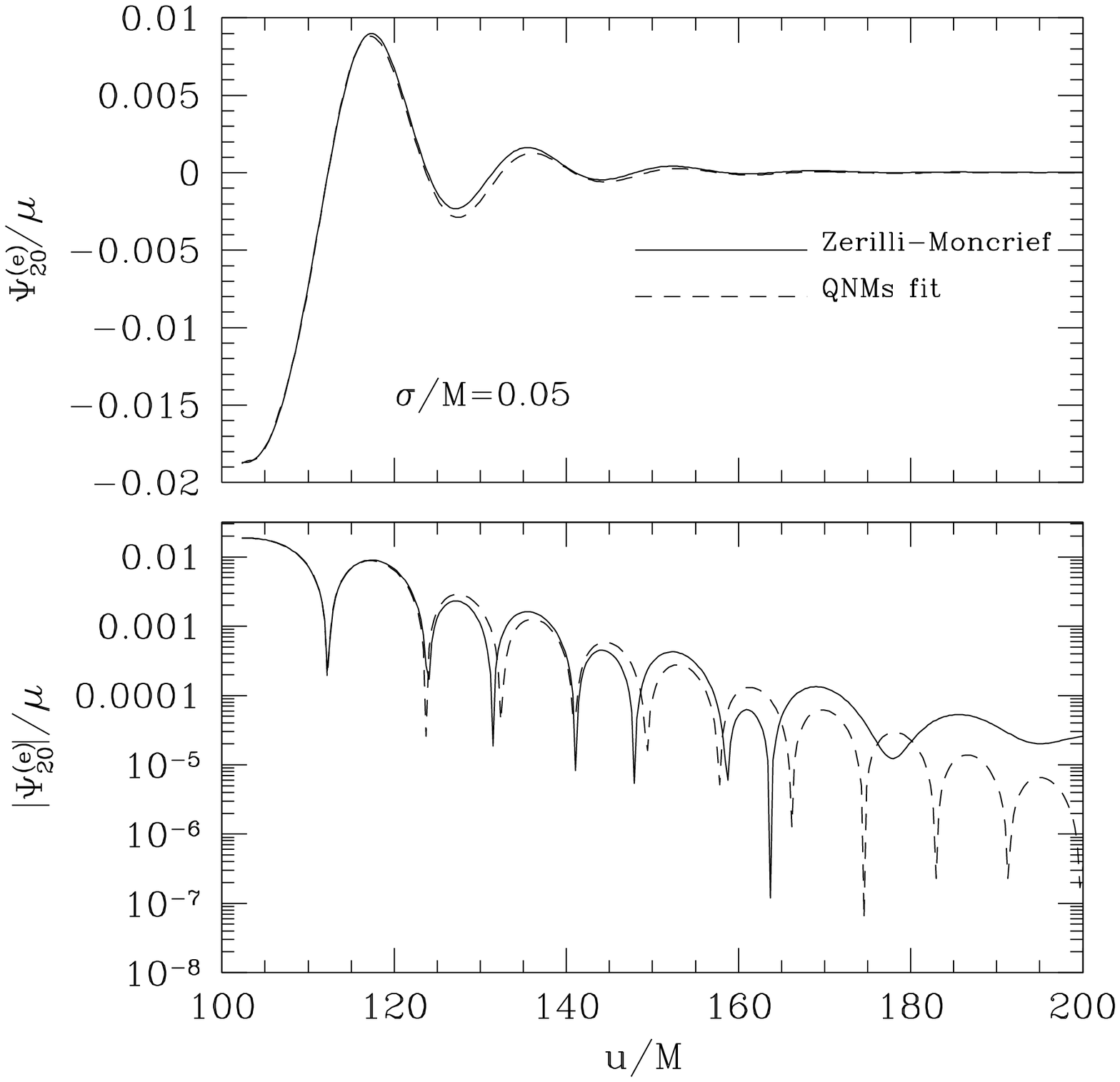}
\vspace{0.0 cm}
\caption{\label{fig1}The $\ell =2$ gravitational Zerilli-Moncrief
  function extracted at $r_{\rm obs}=500\,M$ for a dust shell falling
  from $r_0=15\,M$ with $\sigma/M=0.050$. {\it Left panel:} the
  complete waveform.  {\it Right panel:} investigation of the
  presence of the black hole QNMs: the dashed line was obtained
  through a non-linear fit with the first five modes and the
  comparison with the computed waveform is shown on both a linear and
  a logarithmic vertical scale.}
\end{center}
\vspace{-0.5 cm}
\end{figure*}
%-------------------------------------------------

The P\"oschl--Teller potential is shown in the top left panel of
Fig.~\ref{fig_2} together with $V^{(\rm o)}_2$. Although the 
exponential decay of the former is evident for large $r_*$, it is
also apparent that it provides a good approximation to the
Regge-Wheeler potential near the peak. The accuracy of the
approximation was first studied in Ref.~\cite{ferrari84a}, where 
it was pointed out that the QNMs of the P\"oschl-Teller potential 
can be computed analytically. The frequencies found for the 
lower modes agree within a few percent with those obtained using 
the true curvature potential $V^{(\rm o)}_{\ell}$, computed
numerically in~\cite{Leaver}.

We solve Eq.~(\ref{wave_equation}) using a Lax-Wendroff method
on an evenly spaced $r_*$ grid with $\Delta r_*=0.1\,M$ and with
initial data given by a Gaussian pulse $\Psi_2 =
N\exp\left[-(r-r_{\rm c})^2/\sigma^2\right]$ where $r_{\rm c}$ is the
initial position of the pulse and $N$ a normalization constant. The
initial pulse is considered to be purely ingoing (i.e., with
$\de_t\Psi_2= \de_{r_*}\Psi_2$). We compare the signal
extracted at $r_{\rm obs}=200\,M$ for different values of $\sigma$
using either the Regge-Wheeler or the P\"oschl-Teller potential. This
is summarized in Fig.~\ref{fig_2}, which shows the waveforms obtained
for different values of $r_c$ and $\sigma$ and expressed in the
retarded time, $u=t-r_*^{\rm obs}$. Note that the waveforms from the
P\"oschl-Teller have been shifted in time so as to overlap the maxima 
and the minima of the ringdown phase of the Regge-Wheeler potential.
This is necessary because the scattering of the pulse starts earlier 
for the Regge-Wheeler potential than for the P\"oschl-Teller one.

Let us focus first on the top right panel of Fig.~\ref{fig_2},
which refers to a very narrow pulse ($\sigma=\,M$) initially located at
$r_{\rm c}=100\,M$ and shows with a solid and a dashed line the
waveforms obtained with the Regge-Wheeler and P\"oschl-Teller
potentials, respectively. Clearly, the ringing is very similar in the
two cases both in the wavelength and in the amplitude, but differences
become apparent after $u/M\simeq 180$, when the tail term of the
Regge-Wheeler potential becomes dominant and the gravitational-wave
signal is driven by the backscatter due to the $r^{-2}$
decay and asymptotes the expected late-time decay according to Price's 
law~\cite{price72}. Since the P\"oschl-Teller potential decays exponentially, 
no tail effects are found and the signal is still given by a
superposition of exponentially damped harmonic oscillations. Smaller
differences between the waveforms are however present also in the
early part of the waveform (cf., the inset of the top right
panel of Fig.~\ref{fig_2}) and are, again, due to the fact that
the interaction of the perturbation with the Regge-Wheeler potential
starts ``earlier'' (i.e., at larger radii) than in the case
of the P\"oschl-Teller potential.

The two bottom panels of Fig.~\ref{fig_2} refer instead to a
wave-packet that is initially closer to the black hole (i.e.,
$r_{\rm c}=50\,M$) and show the impact on the waveforms of an increasing
width of the wave-packet. Most notably, the two panels show that as
$\sigma$ is progressively increased, the effects due to the slow-decay
of the Regge-Wheeler potential become progressively more pronounced,
with the ringdown lasting progressively less and with the backscattering
being correspondingly anticipated. For example, when $\sigma=9.5\,M$
(left-bottom panel of Fig.~\ref{fig_2}) and despite the
fundamental mode is still recognizable, other oscillations are present
already after $u/M \sim 70$, while for $\sigma=11.5\,M$ (right-bottom 
panel) the global waveform is essentially overwhelmed by backscattering, the QNMs are
almost absent and the signal is dominated by the ``tail''. An
additional increase in the pulse width would make the QNMs disappear
completely. Clearly, this behavior is not present in the case of a
scattering off a P\"oschl-Teller potential, which is much less
sensitive to the finite-size of the perturbation as long as it 
is smaller than the scale-height set by the exponential decay.

Although just a toy-problem, this simple comparison of the black-hole
response when modeled with the P\"oschl-Teller potential is very
useful to clarify that in the case of general initial data with a
finite size, the shape of the curvature potential strongly affects the
gravitational waveform. In the next sections we will refine this
finding by considering more realistic perturbations such as those
produced by the accretion of extended distributions of matter.

%%%%%%%%%%%%%%%%%%%%%%%%%%%%%%%%%%%%%%%%%%%%%%%%%%%%%%
\subsection{Quadrupolar shells of dust}
\label{shells}
%%%%%%%%%%%%%%%%%%%%%%%%%%%%%%%%%%%%%%%%%%%%%%%%%%%%%%

Following the analysis presented in Paper I, we here consider a number
of aspects of gravitational-wave emission resulting from the accretion
of quadrupolar shells of dust which were not investigated in detail
before. More specifically, we will here discuss: {\emph i)} a more
detailed analysis of the black hole ringdown phase, focusing on black
hole QNMs and on backscattering effects related to the non-exponential
decay of the black hole potential; \emph{ii}) the mechanism
responsible for the production of the interference fringes in the
energy spectra; \emph{iii}) the effect of using conformally-flat
initial data; \emph{iv}) a more detailed analysis of the energy
released in gravitational waves.

As in Paper I, the rest-mass density is parameterized as
\begin{align}
\label{shell}
\rho = {\bar \rho}+\rho_{0}\exp[-(r-r_0)^2/\sigma^2]\sin^2\vartheta \ ,
\end{align}
where $r_0$ is the initial position of the ``center'' of the shell and
$\sigma$ controls its compactness. The background rest-mass density
${\bar \rho}$ is chosen to be very small (i.e., $\sim
10^{-22}$) to simulate the vacuum outside the black hole, while the
normalization constant $\rho_0$ is obtained from the condition that
the volume integral of Eq.~(\ref{shell}) gives $\mu$, the total mass
of the shell, which we choose to be $\mu=0.01M$. 
As we mentioned in Sec.~\ref{sbsc:id} (and Paper I), the initial 
profile for $\Psi^{(\rm e)}_{20}$ is obtained  
after solving the Hamiltonian constraint for $k_{20}$ with $\beta=0$.
Furthermore, to minimize the impact of surious radiation and produce 
initial data that is a ``almost'' time-symmetric, the shell
is kept frozen at $r_0$ (i.e., the hidrodynamics equations are not
evolved) up until the spurious initial gravitational-wave pulse leaves 
the numerical grid. We shall show at the end of this Section how 
waveforms and energy spectra change when ``genuine'' time-symmetric 
and conformally flat initial data are implemented.
%

%-------------------- FIG 4: ENERGY SPECTRA -----------------------------
\begin{figure*}[t]
\begin{center}
\vspace{-0.5 cm}
\includegraphics[width=8.25 cm]{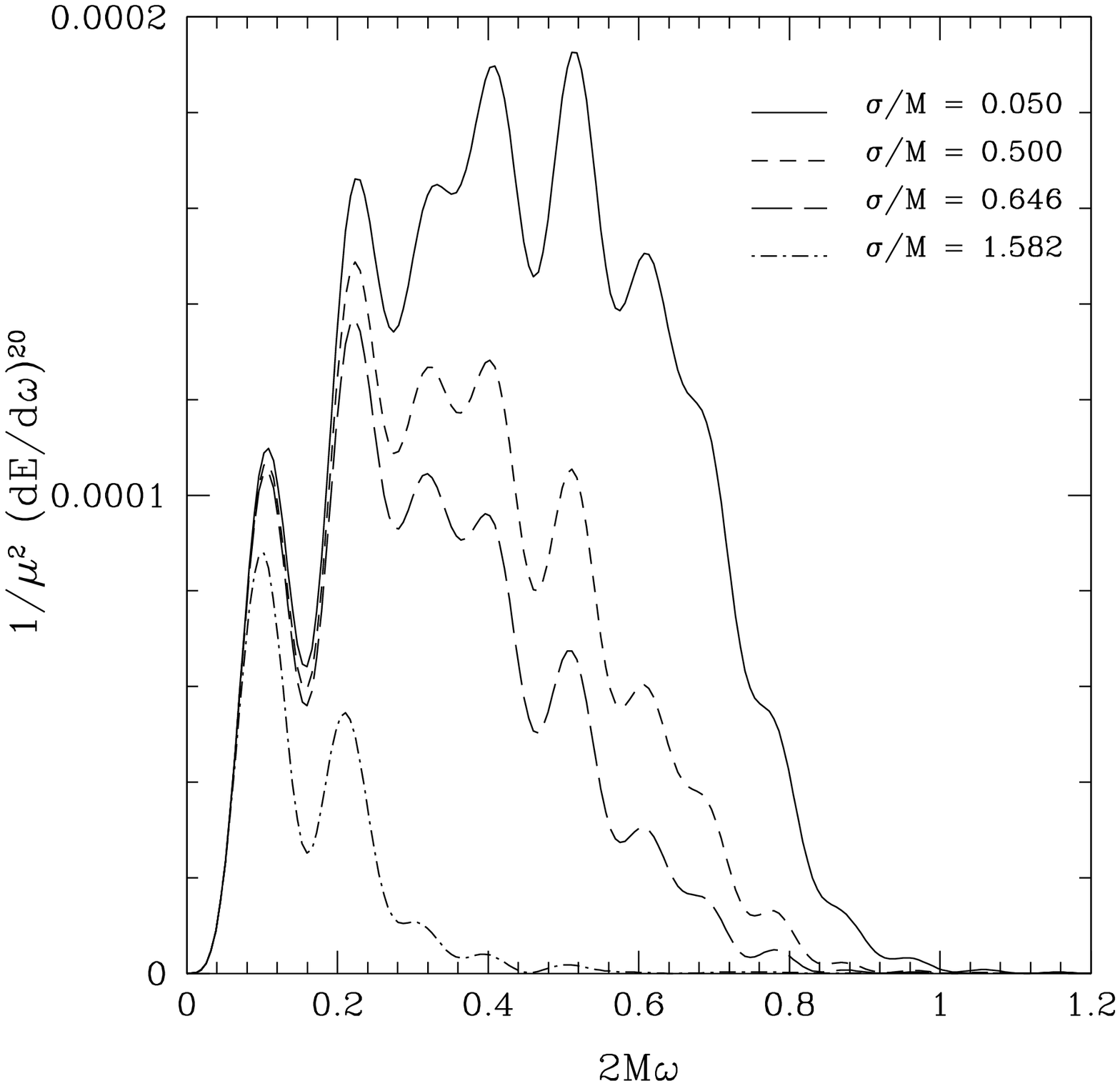}
\hspace{0.5 cm}
\includegraphics[width=8.25 cm]{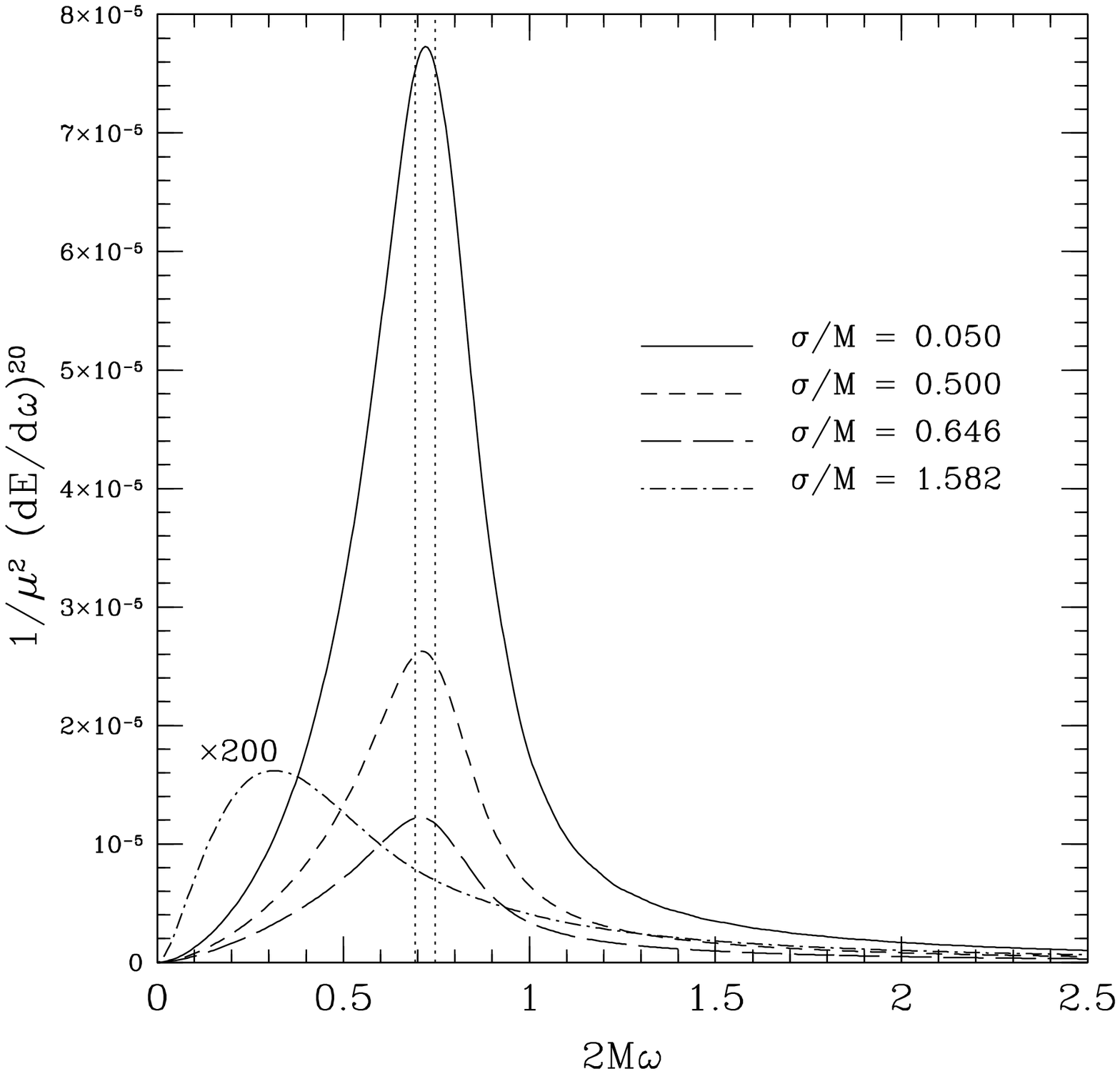}
\vspace{0.0 cm}
\caption{\label{fig3} {\it Left panel:} Energy spectra and their
  typical interference fringes as computed from the waveforms produced
  by shells infalling from $r_0=15M$. {\it Right panel:} The same as in 
  the left panel but only for the ringdown phase, showing the excitation 
  of the $\ell=2$ black hole QNMs. See text for details.}
\end{center}
\vspace{-0.5 cm}
\end{figure*}

%---------------------------- FIG 5: waveforms versus r0------------
\begin{figure*}[t]
\begin{center}
\vspace{-0.5 cm}
\includegraphics[width=8.25 cm]{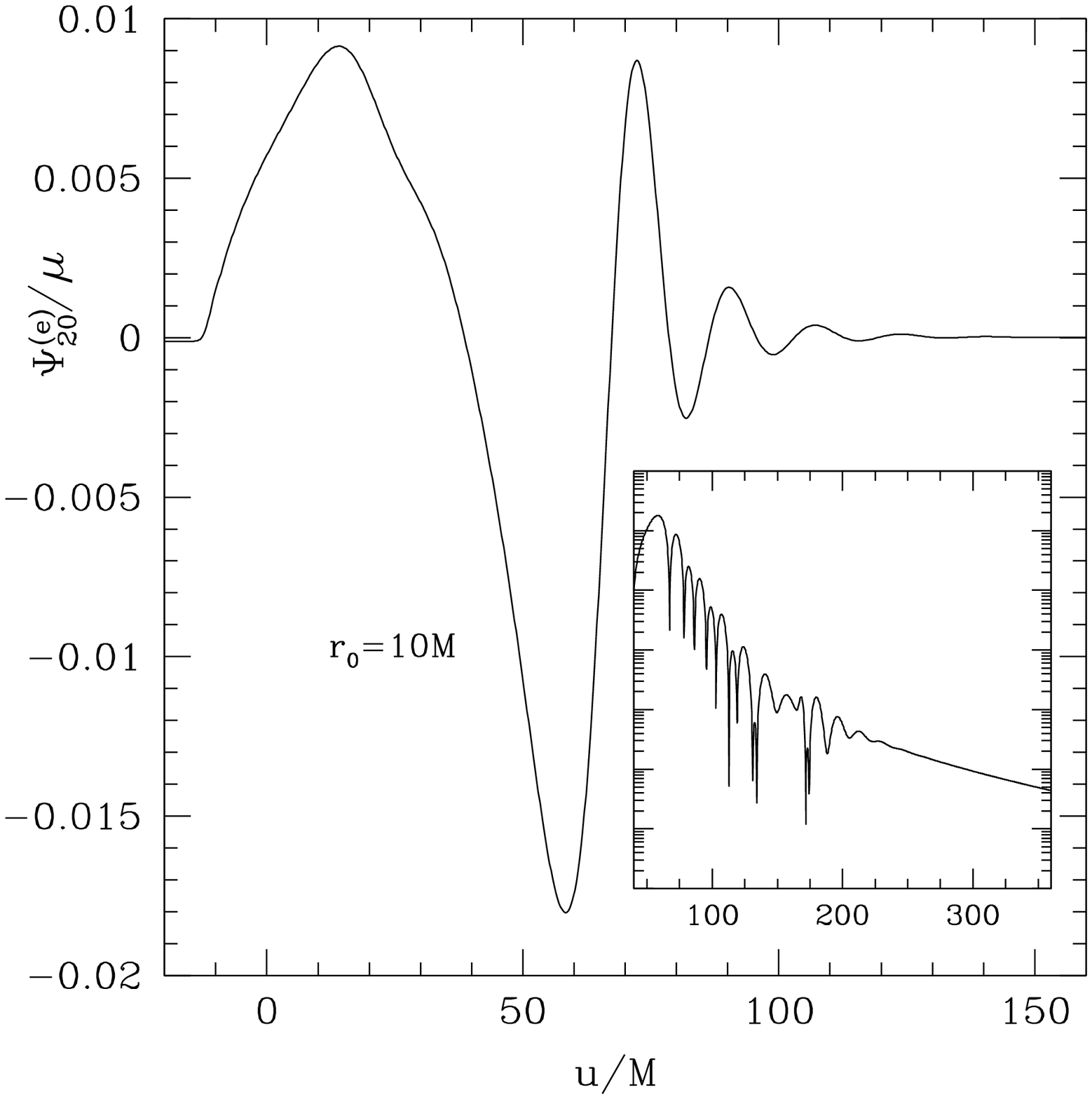}
\hspace{0.5 cm}
\includegraphics[width=8.25 cm]{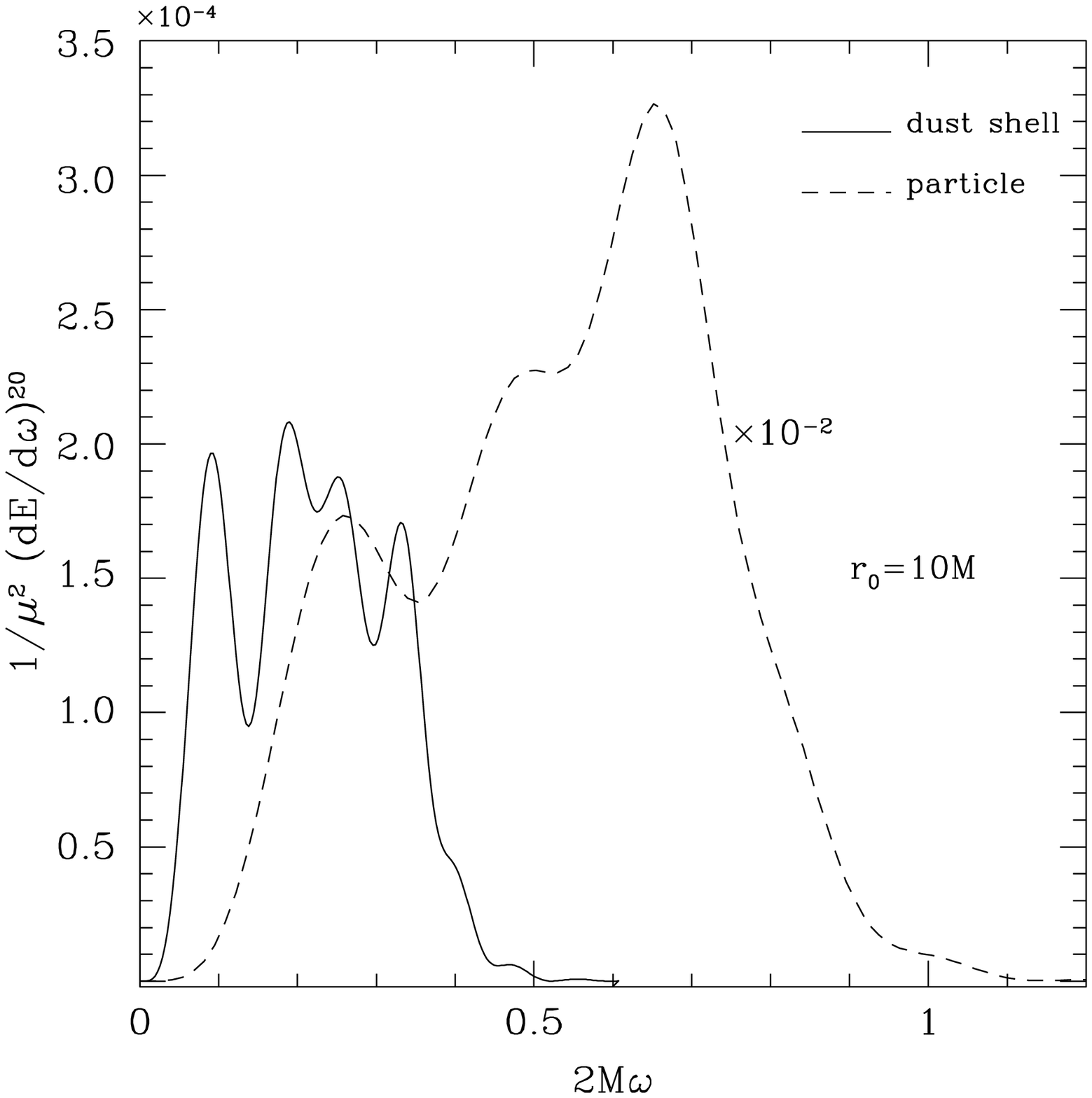}\\
\vspace{-0.5 cm}
\includegraphics[width=8.25 cm]{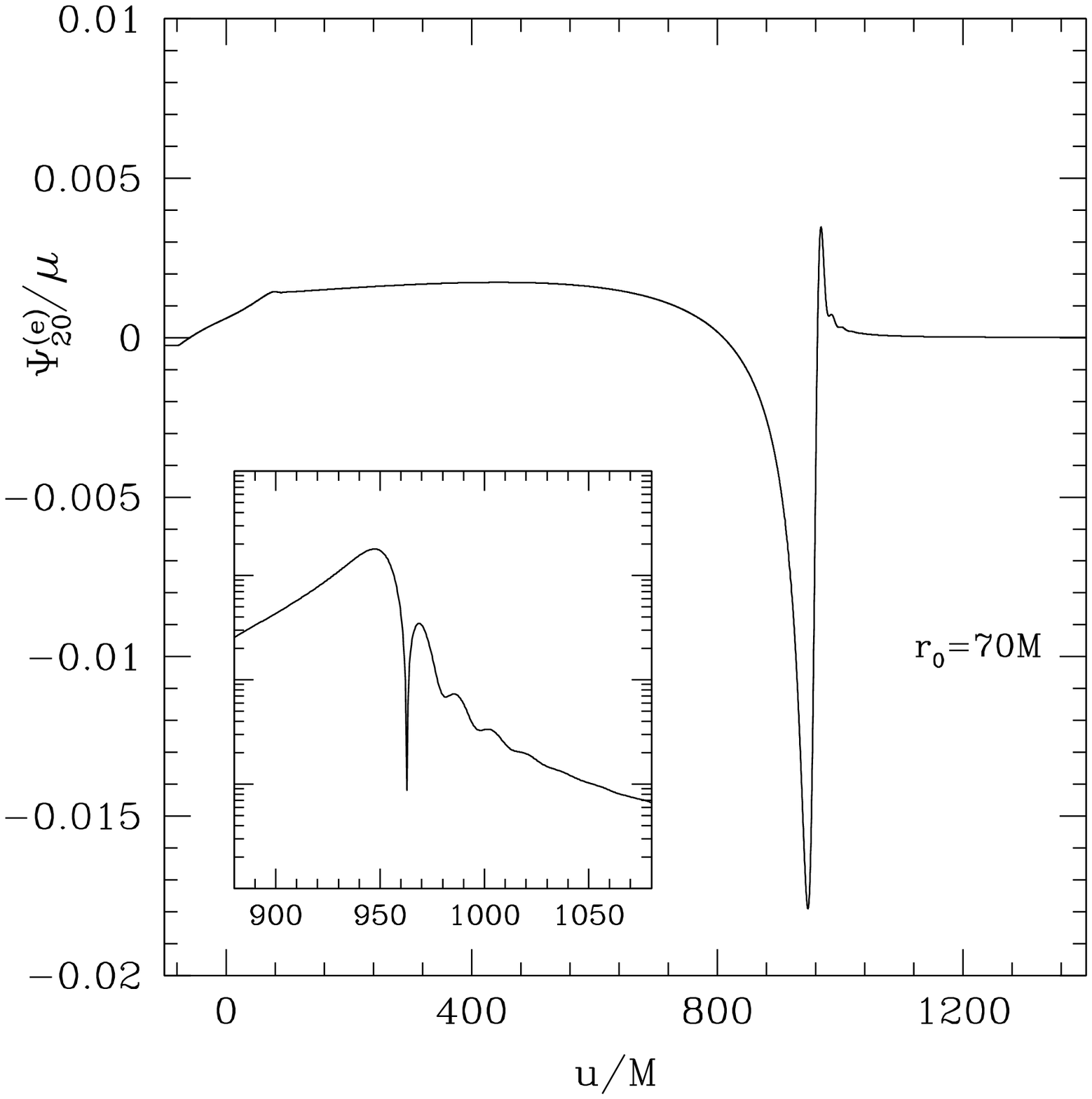}
\hspace{0.5 cm}
\includegraphics[width=8.25 cm]{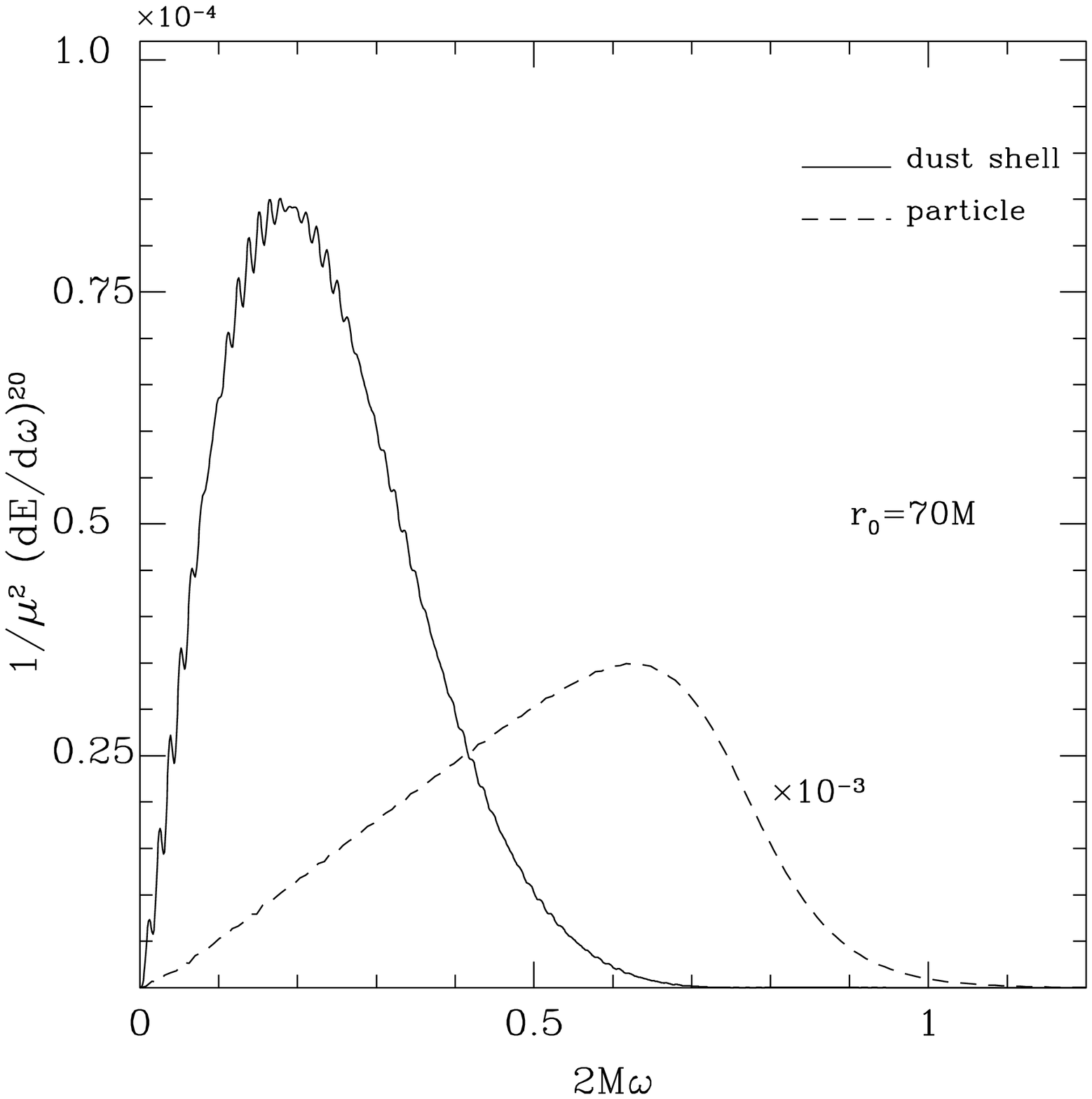}
\vspace{0.0 cm}
\caption{\label{fig5} Infalling shells of dust: waveforms ({\it
    left panels}) and energy spectra ({\it right panels}) for two
  representative values of $r_0$ with $\sigma/M=0.050$. For the
  sake of comparison, the right panels show in dashed lines the
  $\ell=2$ energy spectra for a particle plunging radially onto the
  black hole (rescaled by a convenient factor).}
\end{center}
\vspace{-0.5 cm}
\end{figure*}
%------------------------------------------------------------------

We start by considering the plunge from $r_0=15M$ with
$\sigma/M=0.050$ (this particular model was discussed also in Paper
I) and whose complete waveform is shown in the left panel of
Fig.~\ref{fig1}. As expected, the gravitational-wave signal contains
signatures of both the black hole QNMs and of the backscattering
effects. The first ones are triggered when the maximum of the
rest-mass density crosses the peak of the potential at $r \sim 3M$
yielding the largest contribution to the waveform at $u/M\simeq 112$.
We have verified the presence of the black hole QNMs with complex
frequency $\omega_i$ (with $i=1,2,...$) by fitting the waveform 
from this time onward with a superposition of modes of the form 
$\Psi^{(\rm  e)}=\sum_i\left[\alpha_i \cos(\Re(\omega_i)t)
  +\beta_i\sin(\Re(\omega_i)t) \right]\exp[\Im(\omega_i)t]$, where the
free coefficients $\alpha_i$ and $\beta_i$ are real. The right panel of
Fig.~\ref{fig1} reports the result of such fit obtained using the first
five QNMs, and which is essential to properly capture the early-time
part of the waveform. (Adding higher modes does not
improve the fit, while considering less modes is not enough to
reproduce accurately the waveform.) 
When analyzed on a logarithmic scale (see bottom part of the right 
panel of Fig.~\ref{fig1}), it becomes evident that after $u/M\sim 130$ 
the backscattering effects become dominant and the waveform is no 
longer a simple superposition of exponentially decaying modes.

Additional information on the black hole response comes from the study
of the energy spectra for shells of different initial width $\sigma$
and starting from $r_0=15M$. This is summarized in Fig.~\ref{fig3},
whose left panel shows the energy spectra of the complete waveforms
and the characteristic interference patterns already introduced in
Paper I. Clearly, the interference effects at higher frequencies
(i.e., for $2M\omega \gtrsim 0.2$) as well as the efficiency
in the emission of energy via gravitational waves are increased as the
compactness of the shell is increased, progressively tending 
to what is expected for a point like particle (cf.  the top-right
panel in Fig.~\ref{fig5}). Equally clear is that the energy spectra 
do not show any peculiar behavior around the QNMs frequencies
(i.e., for $2M\omega \gtrsim 0.35$) but also that the
behavior of the spectra at low frequencies does not change
significantly with the matter compactness.

As argued in Paper I and in Refs.~\cite{lousto97a,martel01},
interference fringes naturally arise in the energy spectrum generated by
the superposition of two waveforms separated by a certain time lag
$T$. Indeed, given two time series $A_1(t)$ and $A_2(t)={\cal C} A_1(t+T)$,
the power spectral density resulting from their superposition is given
by
\begin{equation}
\label{modulation}
\frac{dE}{d\omega}\propto
|\tilde{A}_1(\omega)|^2\left[1+{\cal C}^2+2{\cal C}\cos(\omega T)\right] \ , 
\end{equation}
where $\tilde{A}_1(\omega)$ is the Fourier transform of $A_1(t)$. As a
result, such a spectral density will have peaks with a constant
spacing given by $\Delta\omega =2\pi/T$. If ${\cal C}=1$, the 
minima occur at $dE/d\omega=0$ and the spectrum is said to 
have $100\%$ frequency modulations. The modulation is 
always smaller for generic values of ${\cal C}$.

The energy spectra displayed in the left panel of Fig.~\ref{fig3} can be
explained within this general picture. The series of fringes is
mainly determined by the interference of two bursts of radiation with 
different amplitude and separated in time by the
scattering off the curvature potential at different times. In the case
of a shell of width $\sigma/M=0.050$, the two main pulses responsible
for this modulation can be distinguished in the waveform shown in
the left panel of Fig.~\ref{fig1}. These correspond to the peak
produced by the initial motion of $\Psi^{(\rm e)}_{20}$ and appearing at
very early times (i.e., at $u/M\sim 20$), and to the one
emitted when the bulk of accreting matter crosses the peak of the Zerilli
potential (i.e., around $u/M\sim 112$).\footnote{Note that
because $\Psi^{(\rm e)}$ is initially nonzero close to the peak of
the Zerilli potential, some backscattering is also possible before
the bulk of the shell reaches the peak of the potential itself.
This explains the small oscillation in the waveform at $u/M\sim 20$.} 
The time-lag between the two pulses is $\Delta u/M \sim 92$, which gives
a separation between the peaks of the spectrum $2M\Delta\omega\sim 0.07$. 
This value is in good agreement (given the difficulty in unambiguously 
catch the time when the ringdown starts) with the $2M\Delta\omega\sim 0.1$ 
that can be read off from the solid line in the left panel of Fig.~\ref{fig3}.
  
The right panel of Fig.~\ref{fig3}, on the other hand, shows the
spectra obtained by performing a Fourier transform of the signal in
the ``ringdown'' phase only. The two dotted vertical lines indicate
the $n=1$ (fundamental) and $n=2$ (first overtone) QNM frequencies 
of the black hole. It is interesting to note that with the
exception of very wide shells (i.e., with $\sigma/M=1.582$)
which are not able to excite cleanly the QNMs, the maximum of the
spectra always lies between the two lines, suggesting the 
presence of the two modes in the waveform.

Another parameter influencing the energy spectra is the initial
location of the shells $r_0$. As mentioned above, the largest part of
the gravitational-wave signal is emitted when the ``center'' of the
shell reaches the peak of the Zerilli potential and this will clearly
depend on the initial position of the shell. Shells that start further
away will have longer infalling times, larger separations between
the first and second peaks in the waveforms, and thus smaller
separation in the interference fringes of the energy spectra.

In order to reproduce this dependence we have performed a number of
simulations for shells of fixed initial width ($\sigma/M=0.050$) 
accreting from different initial locations $r_0$. Two representative and
extreme cases for $r_0=10M$ and $r_0 = 70M$ are shown in
Fig.~\ref{fig5}, whose left panels display the waveforms, while the
right ones the energy spectra. Overall, these plots show a number of
interesting and general properties of the excitation of black hole
oscillations through accreting matter. Firstly, as the infalling time
is increased, the ringdown phase is progressively dominated by the
non-oscillatory tail, essentially as a result of the ``spreading'' of
the shell induced by the tidal field.  This behavior is indeed
consistent with the analysis carried out in Ref.~\cite{berti06a} and
which pointed out that the excitation of the black hole QNMs through a
continuous flux of infalling particles is made difficult by the
presence of destructive interferences effects. 

Secondly, the efficiency in gravitational-wave emission decreases for
shells falling from larger distances and is significantly smaller than
for point-like particles. This can be easily appreciated  in
the right panels of Fig.~\ref{fig5}, when comparing energy spectra 
associated with shells (solid lines) with the corresponding 
$\ell=2$ energy spectra of a particle plunging radially from the same 
initial position (dashed line)\footnote{The interested reader is
addressed to Paper I and to Ref.~\cite{NDT} for details about the 
handling of a point-like particle in the current framework.}. 
Note that the efficiency is in this case of at least a couple of 
orders of magnitude larger (the data have been properly rescaled 
to aid the comparison). 

\begin{figure}[t]
\begin{center}
\vspace{-0.5 cm}
\includegraphics[width=8.25 cm]{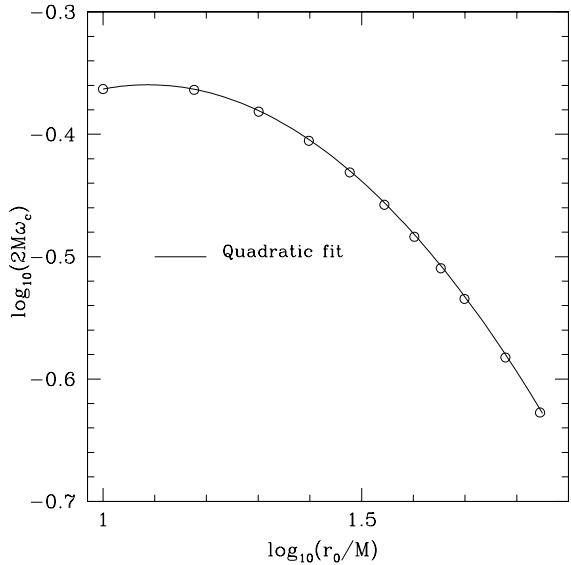}
\vspace{0.0 cm}
\caption{\label{fig6b} Dependence of the characteristic frequency in
  the energy spectrum on the initial position of the shell.}
\end{center}
\vspace{-0.5 cm}
\end{figure}

Thirdly, the energy emission is progressively peaked toward lower
frequencies as $r_0$ increases. This behavior can be made more 
quantitative by defining the \textit{``characteristic frequency''} 
of the energy spectrum as the weighted average
\begin{equation}
\omega_{\rm c} \equiv \dfrac{\int \omega (dE^{20}/d\omega)d\omega} {\int(dE^{20}/d\omega) d\omega} \ ,
\end{equation}
and by computing how this characteristic frequency changes along a
sequence of shells of fixed initial compactness and falling from
increasingly larger radii. A summary of this dependence is offered in
Fig.~\ref{fig6b} for shells with $\sigma/M=0.158$. By following a 
phenomenological approach, if one defines $x=\log_{10}(r_0/M)$ 
and $y=\log_{10}(2M\omega_c)$ the figure shows that the data can be 
very well fitted by means of a quadratic law like $y=ax^2+bx+c$ with
$a=-0.461$, $b=1.002$ and $c=-0.904$. The interpretation of this 
relation is still unclear but certainly deserves further attention.

%----------- FIG 6: contribution of GWs ID  -------------
%-----------------------------------------------------

\begin{figure}[t]
\begin{center}
\vspace{-0.5 cm}
\includegraphics[width=8.25 cm]{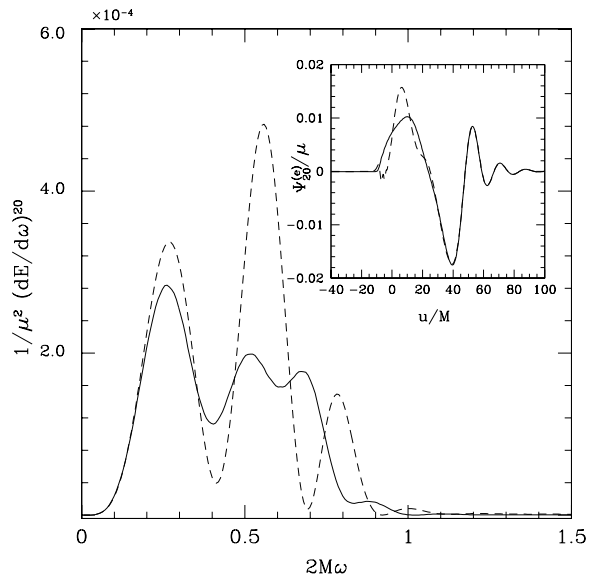} 
\vspace{0.0 cm}
\caption{\label{fig7}Conformally flat initial data and spurious bursts
  of radiation: energy spectra obtained by solving the Hamiltonian constraint 
  with $\beta=0$ only (dashed line) and when the initial gravitational-wave 
  pulse is additionally eliminated (solid line). The shell has $\sigma/M=0.158$, 
  it falls from $r_0=7.5M$ and the signal is extracted at $r_{\mathrm{obs}}=500M$. 
  The inset shows the corresponding initial part of the waveform.}
\end{center}
\vspace{-0.5 cm}
\end{figure}

 As we anticipated at the beginning of this section, an important 
comment is worth making on the influence of the initial amount of 
gravitational radiation on the waveforms and on the energy spectra. 
Two different approaches are possible in this respect and it is 
important to bear in mind that 
they do not yield identical results. The first one 
\footnote{This apprach has  been systematically adopted in 
Refs.~\cite{lousto97a,martel01} for a particle plunging radially from 
a finite distance}
consists in selecting a time-symmetric and conformally flat 
(or non-conformally flat) initial profile of the Zerilli-Moncrief function 
from the solution of Hamiltonian constraint. As we discussed in 
Sec.~\ref{sbsc:id}, this does not prevent that spurious radiation is 
produced as the evolution starts. The second approach, which has been the 
one adopted here,
consists, after the solution of the constraint, in the removal of the 
spurious burst of gravitational radiation by evolving the perturbation 
equations but {\it not} the perturbing sources. In this way the initial 
radiation is allowed to leave the computational domain and the evolution 
can therefore start self-consistently once the initial, nonstationary part 
of the solution has been removed.
Figure~\ref{fig7} summarizes the impact on the energy spectra and on 
the waveforms of the two approaches for a shell initially at rest at 
$r_0=7.5M$. 
Plotted with the dashed line in the main panel is the energy spectrum
in the case of actual conformally flat initial data involving the solution
of Hamiltonian constraint only; on the other hand, the solid line 
refers to the spectrum one obtains when the initial pulse is allowed
to be radiated away. Note the much stronger modulation present in the 
former case and the corresponding larger variation in the initial part 
of the waveform as shown in the inset. Note that we would have found 
larger differences for smaller $r_0$ and smaller differences for 
larger $r_0$. Furthermore, we note that if
we start with non-conformally flat initial data (that is, throught the 
solution of Hamiltonian constraint with $\beta\neq 0$) we would obtain 
larger differences in the early part of the waveforms and thus larger modulations in the
energy spectrum. This behavior is qualitatively similar to what discussed
in Ref.~\cite{martel01} in the case of particles plunging radially from
finite distance

%----------- FIG 7: contribution of GWs ID  -------------
%-------------------------- Table II: energy versus r0 --------------

%
%-- TABLE III: torus models --------------------------
\begin{table*}[t]
\caption{\label{table4}Stable ($D_0$ and $D_1$) and marginally stable
  ($D_2$) constant angular momentum thick disks orbiting around a
  Schwarzschild black hole of mass $M=2.5M_{\odot}$. From left to
  right, the columns report the name of the model, the number of
  radial and polar gridzones used in the hydrodynamicsl simulations, the disk-to-hole
  mass ratio, the polytropic constant $\kappa$ of the isoentropic EOS
  $p=k\rho^{\gamma}$ with $\gamma=4/3$, the value of the specific
  angular momentum $l$, the position of the cusp $r_{\rm cusp}$ and of
  the center $r_{\rm center}$ of the disk, the rest-mass density at
  the center $\rho_c$, the location of the inner ($r_{\rm in}$) and
  outer ($r_{\rm out}$) disk boundaries, the value of the potential
  barrier $\Delta W$, and the orbital period at the center $t_{\rm
    orb}$.}
\begin{ruledtabular}
\begin{tabular}{cccccccccccccc}
 & Model &  $N_r$  &  $N_\theta$  &  $\mu/M$ & $\kappa$ (cgs) &  $l$  & $r_{\rm cusp}$ &$ r_{\rm center}$  
 & $\rho_{\rm c}$ (cgs) &  $r_{\rm in}$ & $r_{\rm out}$ &  $\Delta W$  &  ${\rm t_{orb}}$ (ms) 
\\
\hline
 & $D_0$ & 300  & 150  &  0.0077  & $2.25\times 10^{13}$  & 3.72 & 5.06 & 7.27  &  $7.31\times 10^{12}$ & 5.26  & 9.50   & $-1\times 10^{-4}$ &  1.51 \\
 & $D_1$ & 300  & 150  &  0.0463  &  $9.00\times 10^{13}$ & 3.80 & 4.57 & 8.35  &  $6.86\times 10^{12}$ & 5.21  &  14.54 &      -0.002        &  1.87  \\
 & $D_2$ & 300  & 150  &  0.0779  & $1.05\times 10^{14}$  & 3.80 & 4.57 & 8.35  &  $8.74\times 10^{12}$ &  4.57 & 15.89  &         0          &  1.87  \\
\end{tabular}
\end{ruledtabular}
\end{table*}
%-------------------------------------------------------

\begin{figure}[t]
\begin{center}
\vspace{-0.5 cm}
\includegraphics[width=8.25 cm]{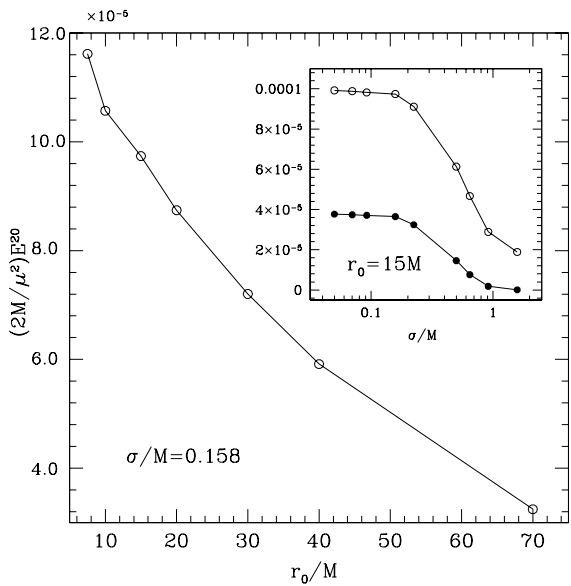}
\vspace{0.0 cm}
\caption{\label{fig8}{\it Main panel:} Total energy radiated in
  gravitational waves ($\l=2$ multipole) as a function of the initial
  location $r_0$ of the center of a shell with $\sigma/M=0.158$. {\it
    Inset:} Total energy versus $\sigma$ for shells falling from
  $r_0=15M$ (open circles).  The filled circles refer to the energy
  contribution coming from from the black hole ringdown only.}
\end{center}
\vspace{-0.5 cm}
\end{figure}

We conclude this section by commenting on the amount of energy
released in gravitational waves as a function of both $r_0$ and
$\sigma$. This is shown in Fig.~\ref{fig8}, which displays in the main
panel the normalized energy emitted in the lowest multipole as a
function of the initial location $r_0$ and for a shell of initial
width $\sigma/M=0.158$. Note that in the case of a shell this is a
monotonically decreasing function of the radial distance, an 
opposite behavior to that seen for accreting particles
where, for conformally flat initial data, the energy has a local
minimum and then increases monotonically with $r_0$~\cite{lousto97a}, 
asymptoting the Davis-Ruffini-Press-Price limit~\cite{DRPP}. 
In addition, the inset in Fig.~\ref{fig8} displays a comparison 
between the energy emitted through the whole waveform (empty circles) 
and the one computed considering the ringdown phase only (filled circles). 
Note that while the black hole ringdown contributes for 
$\sim 30\%-40\%$ of the total energy in the case of small $r_0$, 
this value goes down to $\sim 10\%$ for larger initial distances 
as a result of the progressive loss of compactness experienced by 
the shell during the infall.
%
%--------------------- FIG 7: plot of the total energy --------------
%--------------------------------------------------------------------

%----------------------------------
\subsection{Thick accretion disks}
\label{tori}
%----------------------------------
%
%-------------------------- FIG.8: Equatorial profile of the density ---------
\begin{figure*}[t]
\begin{center}
\vspace{-0.5 cm}
\includegraphics[width=8.25 cm]{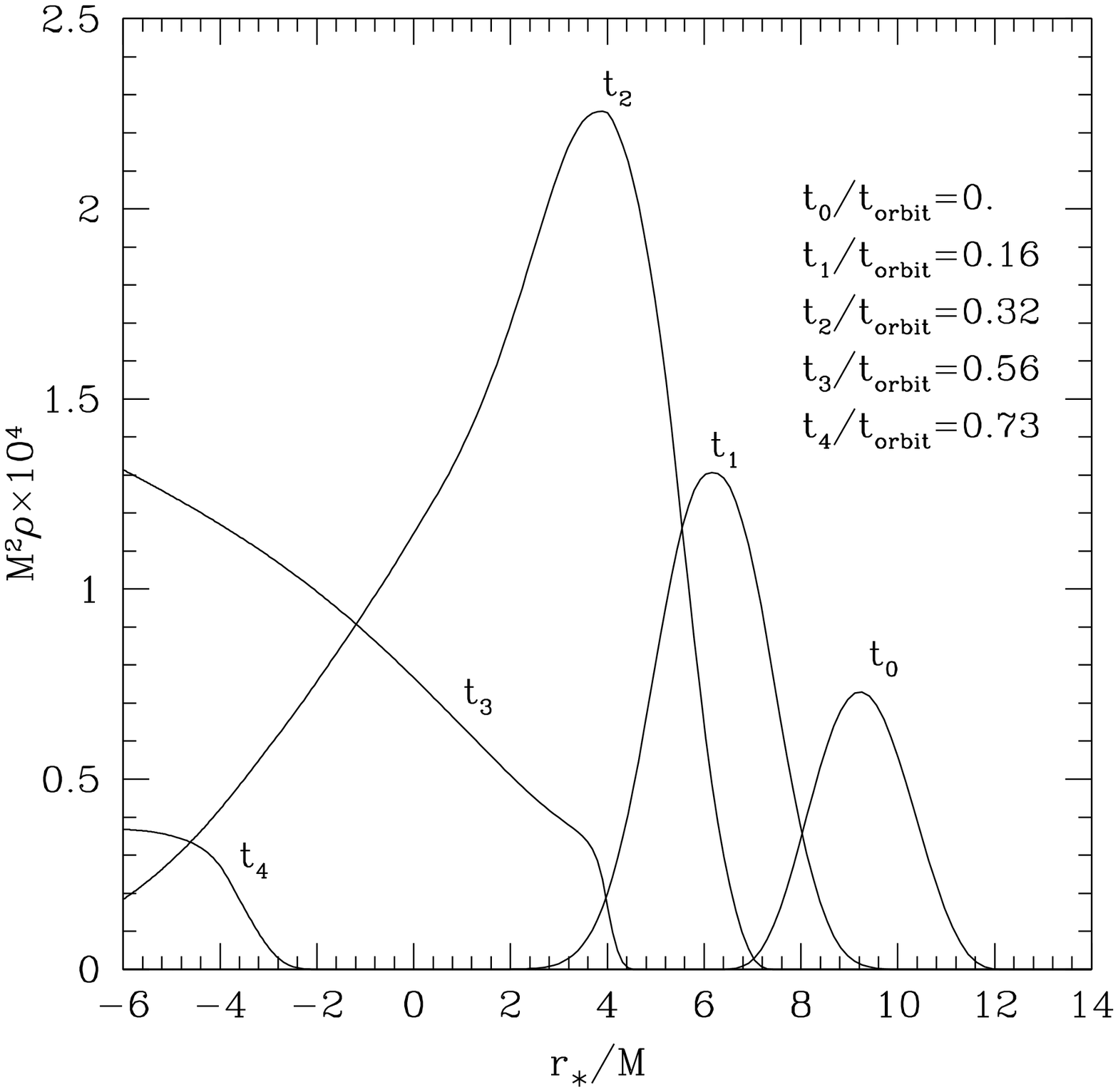} 
\hspace{0.5 cm}
\includegraphics[width=8.25 cm]{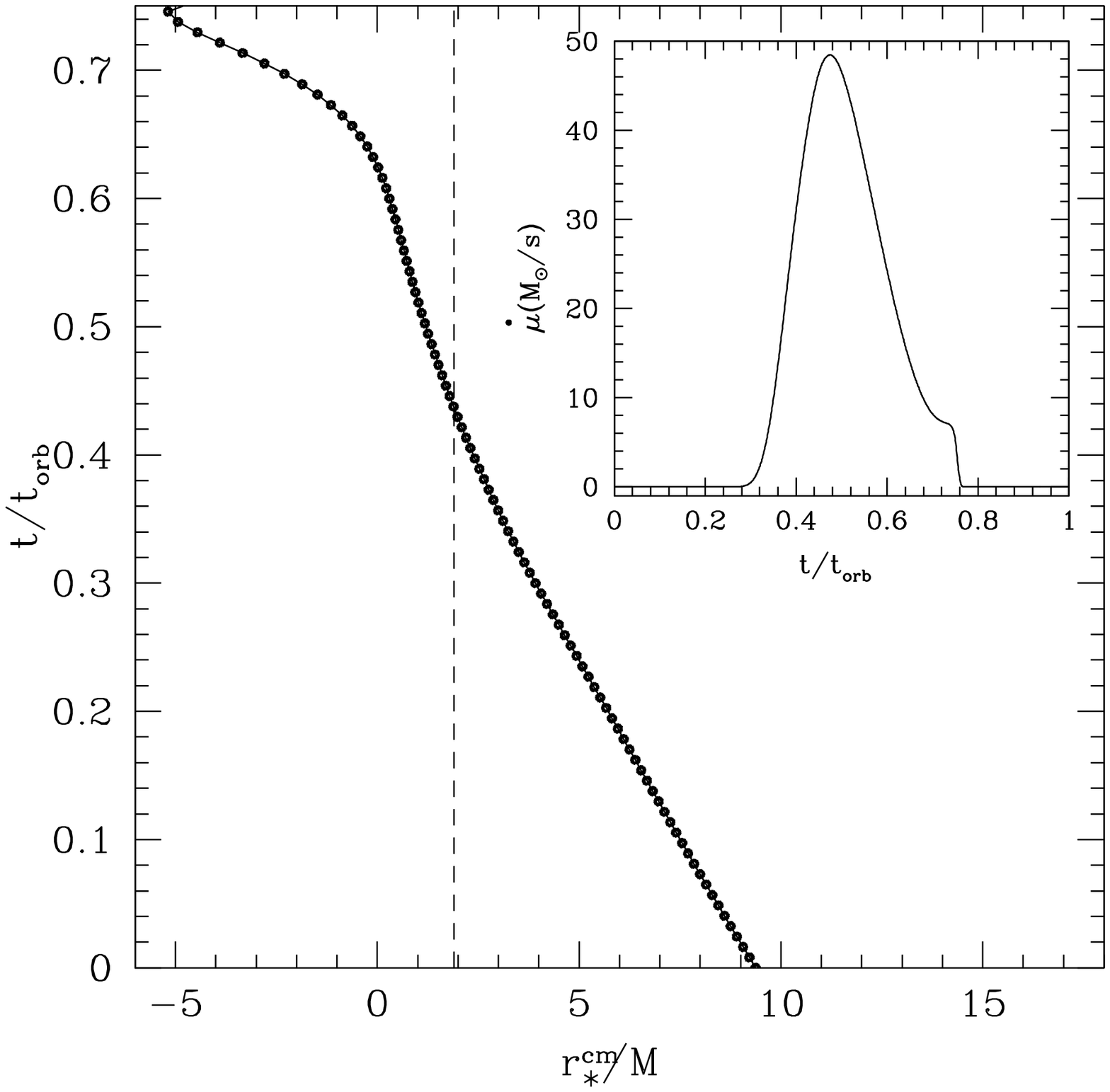} 
\vspace{0.0 cm}
\caption{\label{fig9}Model $D_0$. {\it Left panel:} time evolution of
  the equatorial rest mass density profiles for a velocity
  perturbation $\eta=0.3$. {\it Right panel:} time evolution of the center 
  of mass in the equatorial plane. The disk disappears in the black hole 
  at $t\approx 0.75\, t_{\rm orb}$.}
\end{center}
\vspace{-0.5 cm}
\end{figure*}

While our model with infalling quadrupolar shells allows us to 
capture some of the essential features of the gravitational-wave 
emission from extended matter sources, such as the excitation of 
the black hole QNM ringdown and the presence of interference 
effects in the energy spectra, it nevertheless remains a useful 
toy model.

To improve on this and to examine a scenario which is astrophysically
more realistic, we will now consider the gravitational-wave emission
resulting from the accretion of geometrically thick disks (or tori)
orbiting around a non-rotating black hole. Such systems are believed 
to form in a
variety of different ways, such as during the last stages of
gravitational collapse or in the merger of binary neutron stars. In
addition, if compact enough and undergoing oscillations, these systems
generate gravitational-wave signals within the sensitivity curve of
ground-based interferometers~\cite{zanotti03,zanotti05,nagar05a}.
Hereafter, and as in the case of dust shells, we will assume that
the mass of the torus $\mu$ is much smaller than that of the black
hole.

Detailed descriptions on how to build equilibrium configurations of
barotropic thick disks orbiting black holes are given in
Refs.~\cite{abramovicz78,font02a}, but we here recall that these
objects have traditionally been described as obeying a polytropic
equation of state $p=\kappa\rho^{\gamma}$, with $\gamma=4/3$. Because
in pure orbital motion, the fluid four-velocity is given by
$u^{\alpha}=(u^t,0,0,u^\phi)$ and it describes a non-Keplerian
rotation around the central black hole, with a specific angular
momentum $\ell=-u_{\phi}/u_t$ distribution which is essentially 
unknown.

Table~\ref{table4} lists the main features of the tori considered 
in our simulations, such as the values of the specific angular 
momentum $\ell$ (assumed constant throughout the disk), of the 
potential gap $\Delta W$ between the inner edge of the disk and
the cusp, and of the mass of the disk $\mu$. In order to induce a
non-trivial dynamics in these otherwise stationary disks, the initial
models have to be perturbed in some way, such as by adding a radial velocity
$v_r=\eta v_r^{\rm sph}$, where $\eta$ is a parameter and $v_r^{\rm
  sph}$ is the radial velocity of the spherical stationary atmosphere
surrounding the torus~\cite{zanotti03}. The rest-mass density of the
atmosphere is low enough not to influence the dynamics of the tori
(its density is several orders of magnitude smaller than the maximum
density at the center of the disk).

As in the case of infalling dust shells, the initial data for the
perturbation fields $\Psi^{(\rm e/o)}_{\ell 0}$ is obtained either
through the solution of the Hamiltonian constraint (for even-parity
perturbations) with conformally flat initial data, or assuming
stationarity, i.e., through Eqs.~(\ref{R-W}) with $\de^2_t\Psi^{(\rm
  e/o)}_{\ell 0}=0$. Clearly, in either case this initial data is not
consistent with the hydrodynamical sources once they are
perturbed. This mismatch inevitably introduces a unphysical initial
burst of gravitational radiation that is however easy to distinguish
and remove from the analysis.

We note that the accretion of matter onto the black hole can occur on
different timescales depending on the perturbation and we will
consider here the two limiting situations that could be encountered in
astrophysical scenarios. In the first case we will consider a
perturbation which is large enough to cause a runaway accretion of the
disk onto the black hole; we will refer to this as the
``hypercritical accretion'' scenario (see Sect.~\ref{has}). In the
second case, on the other hand, the perturbation is smaller and it
will induce a series of quasi-periodic oscillations, each accompanied
by an episode of accretion onto the black hole; we will refer to this
as to the ``quasi-period accretion'' scenario (see Sect.~\ref{qpa}).

%----------------------------------------------------------------------
\subsubsection{Hypercritical Accretion}
\label{has}
%----------------------------------------------------------------------

As mentioned above, simulations of hyper-accreting disks can be
performed by simply choosing a sufficiently large initial value of the
radial velocity perturbation. Under this condition, the centrifugal 
barrier cannot counteract the injection of additional kinetic energy 
and the whole disk is ``pushed'' towards the black hole, rapidly 
accreting onto it in less than one orbital period.

Hereafter we will concentrate on the dynamics of model $D_0$, to which
a velocity perturbation with $\eta=0.3$ is added. For this model the
hydrodynamical grid runs from $r_{\rm min}=2.03\,M$ to $r_{\rm max}=16
\,M$ and it is covered by $300\times 150$ grid-points, geometrically
spaced in the coordinate $r$, with $\Delta r_{\rm min}= 3\times
10^{-4}$ and $\Delta r_{\rm max}=0.2$.  The range of the $r_*$ 1D grid
is $r_*\in[-2200\,M, \, 4500\,M]$, to guarantee that the outer
boundary does not influence the slope of the late time power-law tail,
and is covered by$\sim3\times 10^4$ cells.

The left panel of Fig.~\ref{fig9} shows, in terms of the tortoise
coordinate, five snapshots of the evolution of the rest-mass density on the
equatorial plane. Note that as time proceeds, the disk is compressed,
the density increases and matter quickly starts falling onto the black
hole. This continues until the entire disk has been accreted. 
A summary of this is shown in the spacetime diagram in the right panel
of Fig.~\ref{fig9}, which presents the motion of the projection on the
equatorial plane of an effective ``center-of-mass'' of the disk,
defined as
\begin{equation}
r_{\rm cm}\equiv\dfrac{\int \sqrt{g_{rr}} \rho r^2dr}
{\int \sqrt{g_{rr}}\rho rdr } \ ,
\end{equation}
where $g_{rr}=(1-2M/r)^{-1}$ (for convenience the spacetime diagram is
shown using the tortoise radial coordinate $r_*$). Correspondingly,
the inset in Fig.~\ref{fig9} depicts the time evolution of the 
mass accretion rate for model $D_0$.

The vertical dashed line in the right panel of Fig.~\ref{fig9} 
is approximately the location of the peak of the Zerilli
potential, $r_*^{\rm peak}\simeq 1.90\,M$ ($r^{\rm peak}\simeq
3.1\,M$) and can be used to identify $t\simeq 0.44\,t_{\rm orb}$ as the
time at which the center of mass crosses the peak of the
potential. Furthermore, this panel also shows that the center 
of mass leaves the hydrodynamical grid ($r_*^{\rm min}\simeq -6.35\,M$) 
at $t\sim 0.75\,t_{\rm orb}$ and from this time onward, the
gravitational-wave signal is dominated by the QNMs of the 
black hole.

%%%%%%%%%%%%%%%%%%%%%%%%%%%%%%%%%%%%%%%%%%%%%%%%%%%%%%%%%%%
%                            Model D0: waveform
%%%%%%%%%%%%%%%%%%%%%%%%%%%%%%%%%%%%%%%%%%%%%%%%%%%%%%%%%%%
\begin{figure}[t]
\begin{center}
\vspace{-0.5 cm}
\includegraphics[width=8.25 cm]{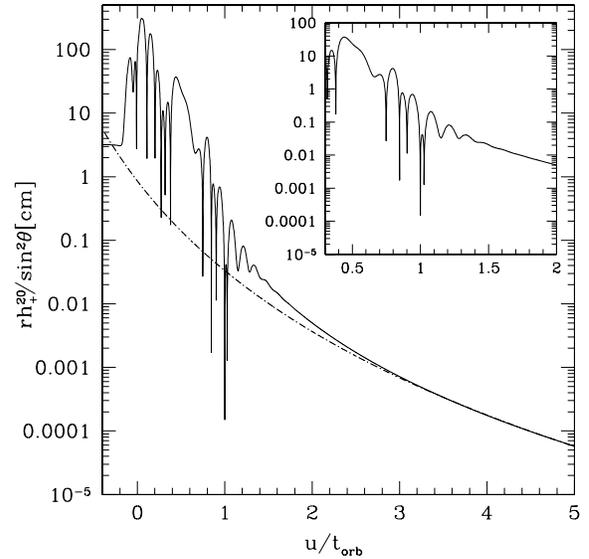}
\vspace{0.0 cm}
\caption{\label{fig10}Model $D_0$: gravitational waveform 
   (extracted at $r_{\rm obs}=200\,M$). The presence of two ringings 
   is evident. The first
  one is related to the gravitational-wave content present in the
  initial profile of the Zerilli-Moncrief equation. The second ringing
  (magnified in the inset) is determined by the object crossing the
  peak of the potential. The dashed line, proportional to $u^{-7}$,
  shows that the late-time behavior of the waveform is properly
  captured.}
\end{center}
\vspace{-0.5 cm}
\end{figure}
%%%%%%%%%%%%%%%%%%%%%%%%%%%%%%%%%%%%%%%%%%%%%%%%%%%%%%%%%%%
The gauge-invariant gravitational waveform produced by the
hypercritical accretion has a dominant $\l=2$ character\footnote{Note
  that because the system is axisymmetric, $\Psi^{(\rm e/\rm
    o)}_{\l 0} = 0$ when $m\neq 0$, and $\Psi^{(\rm e)}_{20} \gg
  \Psi^{(\rm o)}_{20} \approx 0$, $\Psi^{(\rm o)}_{30} \gg \Psi^{(\rm
    e)}_{30} \approx 0$.} 
and is displayed
in a logarithmic scale in Fig.~\ref{fig10} as measured by an observer
located at $r_{\rm obs}=200\,M$. Besides the initial burst of
spurious radiation related to the initial data, and which is no longer
present by $u\simeq 0.2\,t_{\rm orb}$, the waveforms exhibits two main 
features. The first one is determined by the infalling matter as 
its center of mass approaches the maximum of the potential 
at $u\simeq 0.4\, t_{\rm orb}$. The second one, instead, starts at 
$u\simeq 0.7\,t_{\rm orb}$ and represents the ringing resulting from 
the perturbation experienced by the black hole through the rapid 
accretion of the torus (see the mass accretion rate inset in the right 
panel of Fig.~\ref{fig9}). This becomes apparent when comparing the 
features of the waveform in Fig.~\ref{fig10} with the matter dynamics 
in Fig.~\ref{fig9}.
The ringdown phase ends with the usual power-law tail; this is highlighted
by the dot-dashed line in Fig.~\ref{fig10} which is proportional to
$u^{-7}$ and shows that for $u>3.6\,t_{\rm orb}$ the waveform has only
a quadrupolar nature decaying as $\propto u^{-(2\ell+3)}$.
%
%%%%%%%%%%%%%%%%%%%%%%%%%%%%%%%%%%%%%%%%%%%%%%%%%%%%%%%%%%%%%%%%%%%%%%
%                      FIG. 9 : cm of Model D1
%%%%%%%%%%%%%%%%%%%%%%%%%%%%%%%%%%%%%%%%%%%%%%%%%%%%%%%%%%%%%%%%%%%%%%
%
\begin{figure}[t]
\begin{center}
\vspace{-0.5 cm}
\includegraphics[width=8.25 cm]{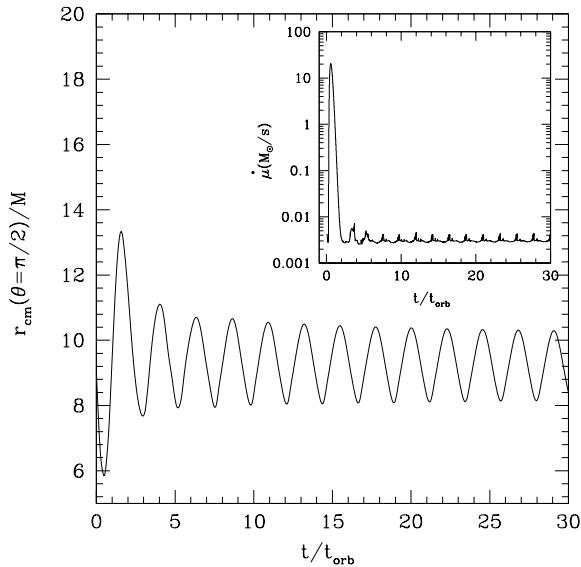}
\vspace{0.0 cm}
\caption{\label{figV_5} Model $D_1$: time evolution of the projection 
  of the center of mass in the equatorial plane. After an initial transient, 
  the dynamics is characterized by constant amplitude quasi-periodic oscillations.
  The inset shows the mass accretion rate.}
\end{center}
\vspace{-0.5 cm}
\end{figure}
%%%%%%%%%%%%%%%%%%%%%%%%%%%%%%%%%%%%%%%%%%%%%%%%%%%%%%%%%%%%%%%%%%%%%%s

In summary, the analysis of the process shows that a very rapid and
hypercritical accretion of a compact distribution of matter onto a
black hole produces gravitational-wave emission that is qualitatively
similar to that of either relatively narrow dust shells plunging from
a large distance or wide dust shells accreting from a small
distance. Furthermore, the gravitational-wave signal shows
oscillations that reflect the excitation of the black hole QNMs
(especially the fundamental one) but these oscillations are in general
so weak that the non-oscillatory tail determined by the long-range
properties of the scattering potential soon dominates the signal.

%%%%%%%%%%%%%%%%%%%%%%%%%%%%%%%%%%%%%%%%%%%%%%%%%%%%%%%%%%%%%%%%%%%%%%
%                   FIG.10: waveforms for Model D1
%%%%%%%%%%%%%%%%%%%%%%%%%%%%%%%%%%%%%%%%%%%%%%%%%%%%%%%%%%%%%%%%%%%%%% 
\begin{figure*}[t]
\begin{center}
\vspace{-0.5 cm}
\includegraphics[width=8.25 cm]{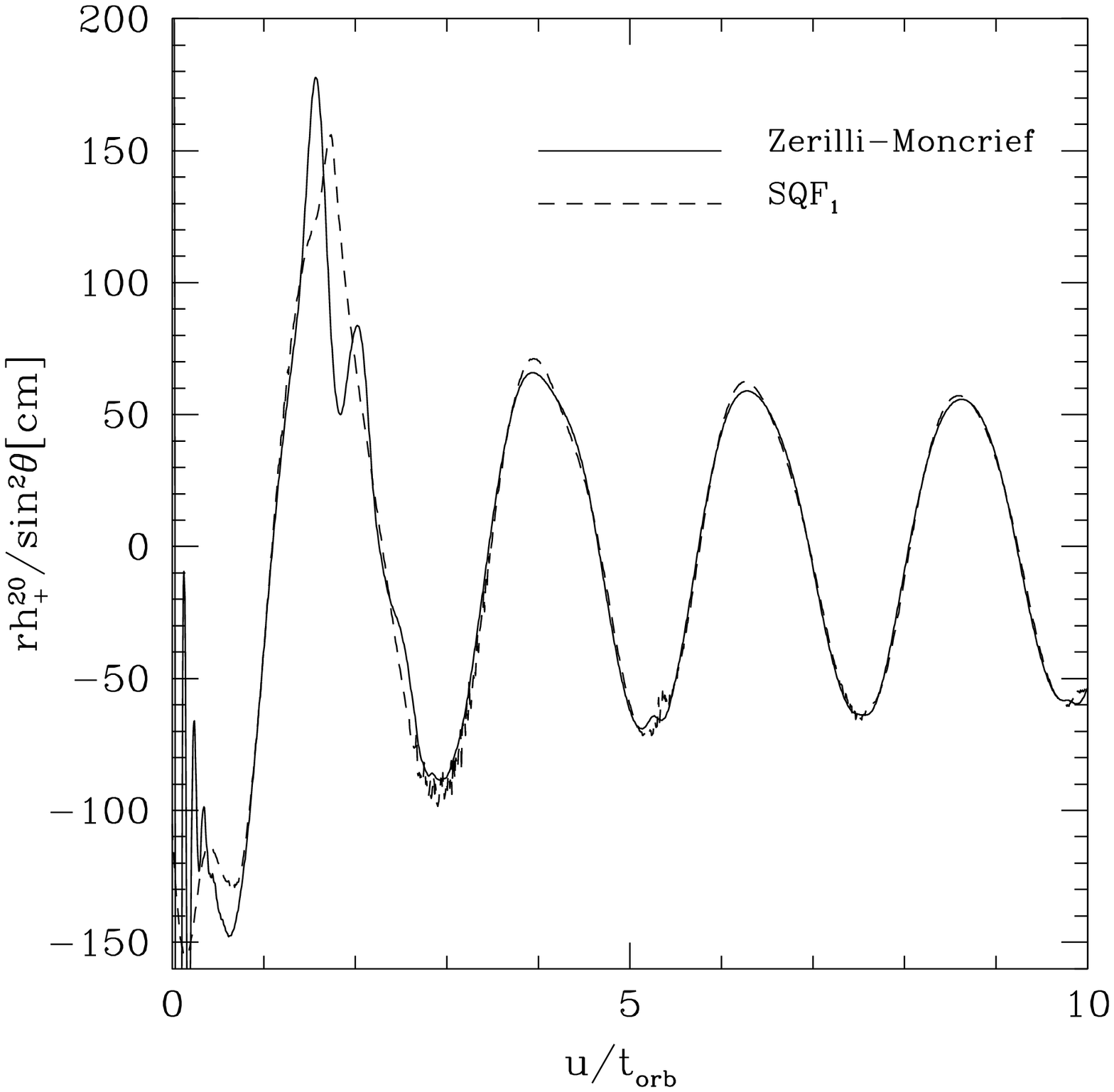} 
\hspace{0.5 cm}
\includegraphics[width=8.25 cm]{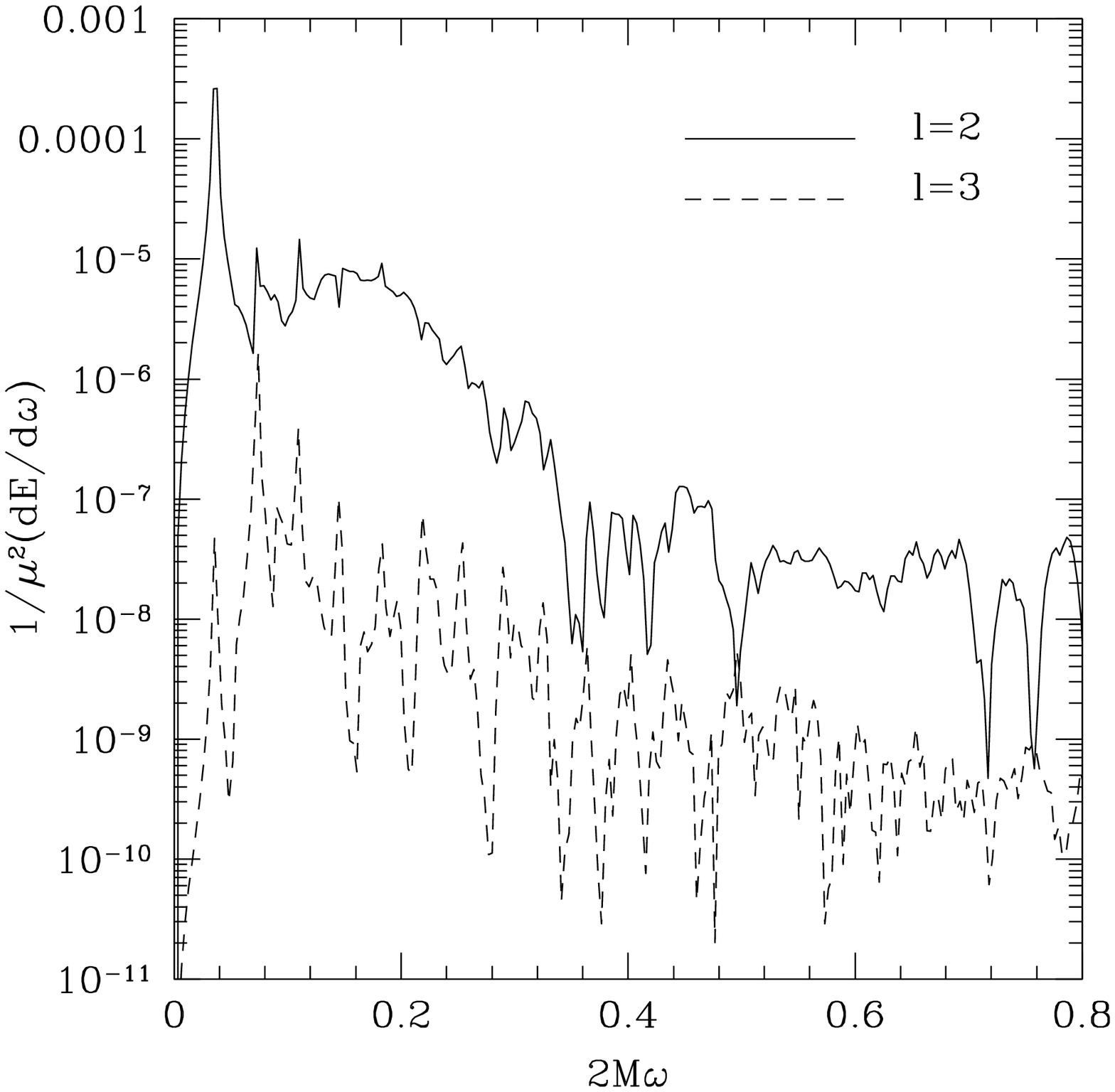}
\vspace{0.0 cm}
\caption{\label{figV_3} Model $D_1$. {\it Left panel}: gravitational 
waveforms (extracted at $r_{\rm obs}=500\,M$) versus retarded time.
$h_+^{20}$ polarization (even-parity) superposed to the waveform
extracted using the SQF$_1$ (dashed line). {\it Right panel}: Energy
spectrum for the $\l=2$ and $\l=3$ (odd-parity) modes. The highly 
damped oscillation in the early part of the $\l=2$ waveform (from $\Psi^{(\rm e)}_{20}$) 
results in a broad-band peak which is not related to the QNMs 
frequencies of the black hole.}
\end{center}
\vspace{-0.5 cm}
\end{figure*}
%%%%%%%%%%%%%%%%%%%%%%%%%%%%%%%%%%%%%%%%%%%%%%%%%%%%%%%%%%%%%%%%%%%%%%
%-------------------------------------
\subsubsection{Quasi-periodic accretion}
\label{qpa}
%-------------------------------------

A phenomenology complementary to the one discussed in the previous
section for a hypercritical accretion can be studied when the
accretion takes place in a quasi-period fashion. Indeed, there are 
at least two different ways of triggering a quasi-periodic accretion 
and we will discuss them separately.

The first possibility has already been discussed in Sect.~\ref{tori}
and consists in adding a perturbation in the radial velocity of the
torus parameterized in terms of an analytic accretion solution. As
a concrete example we will now concentrate on the dynamics of model 
$D_1$, describing a disk with a comparatively high value of the specific
angular momentum, $\l=3.80$, and subject to an initial radial velocity
perturbation with $\eta=0.2$.

Figure~\ref{figV_5} shows the motion of the center of mass on the
equatorial plane, indicating that after an initial transient, a phase
of quasi-periodic, (almost) constant amplitude oscillations follows,
in which the disk periodically approaches the black hole before the
centrifugal barrier pushes it back, past its original position. As the
torus nears the black hole, part of its matter is spilled through the
cusp, resulting in a quasi-periodic accretion of matter onto the
black hole; this is shown by the inset of Fig.~\ref{figV_5} which
reports the evolution of the mass accretion rate (see
Ref.~\cite{zanotti03} for a more detailed discussion). Note that the
accretion is essentially shut-off as the torus moves away from the
black hole.

This quasi-periodic dynamics of the torus is clearly imprinted onto
the gravitational-wave signal and this is shown in the left panel of
Fig.~\ref{figV_3}, where the $\l=2$ gauge-invariant waveform
(extracted at $r_{\rm obs}=500\,M$) is superposed to the signal
extracted using the quadrupole formula (SQF$_1$). Note that the very 
early time part of the gauge-invariant signal has been removed to avoid
influences coming from the burst related to the initial data.  The
waveforms show a small burst at $u\approx t_{\rm orb}$, that
corresponds to the initial accretion of matter, followed by regular
oscillations that mirror the motion of the torus in the potential
well. Some differences between the gauge-invariant and the SQF$_1$
waveforms are recognizable in the initial stages, with the first one
showing highly damped high-frequency oscillations for $u\approx
2\,t_{\rm orb}$ superposed to the main, quasi-periodic signal. Since
this high-frequency pattern is absent in the quadrupole waveform, we
conclude that it originates from curvature backscattering of
gravitational waves emitted in the initial infalling phase.

We also note that the differences between the waveforms computed
through the gauge-invariant formalism and the quadrupole formula using
the quadrupole tensor defined by Eq.~(\ref{sqf1}) become larger if the
initial velocity perturbation is increased. This is simply due to the
fact that the SQF$_1$ is valid in the weak-field and slow-motion
approximation and thus ceases to be accurate when the dynamics of the
oscillations is nonlinear, with the velocities becoming relativistic.
On the other hand, the velocity-dependent terms in the source function
of the Zerilli-Moncrief equation are obviously accurate also in a
relativistic regime.

%%%%%%%%%%%%%%%%%%%%%%%%%%%%%%%%%%%%%%%%%%%%%%%%%%%%%%%%%%%%%%%%%%%%%%%%%%
%                   FIG.12: properties of model D2 
%%%%%%%%%%%%%%%%%%%%%%%%%%%%%%%%%%%%%%%%%%%%%%%%%%%%%%%%%%%%%%%%%%%%%%%%%%
\begin{figure*}[t]
\begin{center}
\vspace{-0.5 cm}
\includegraphics[width=8.25 cm]{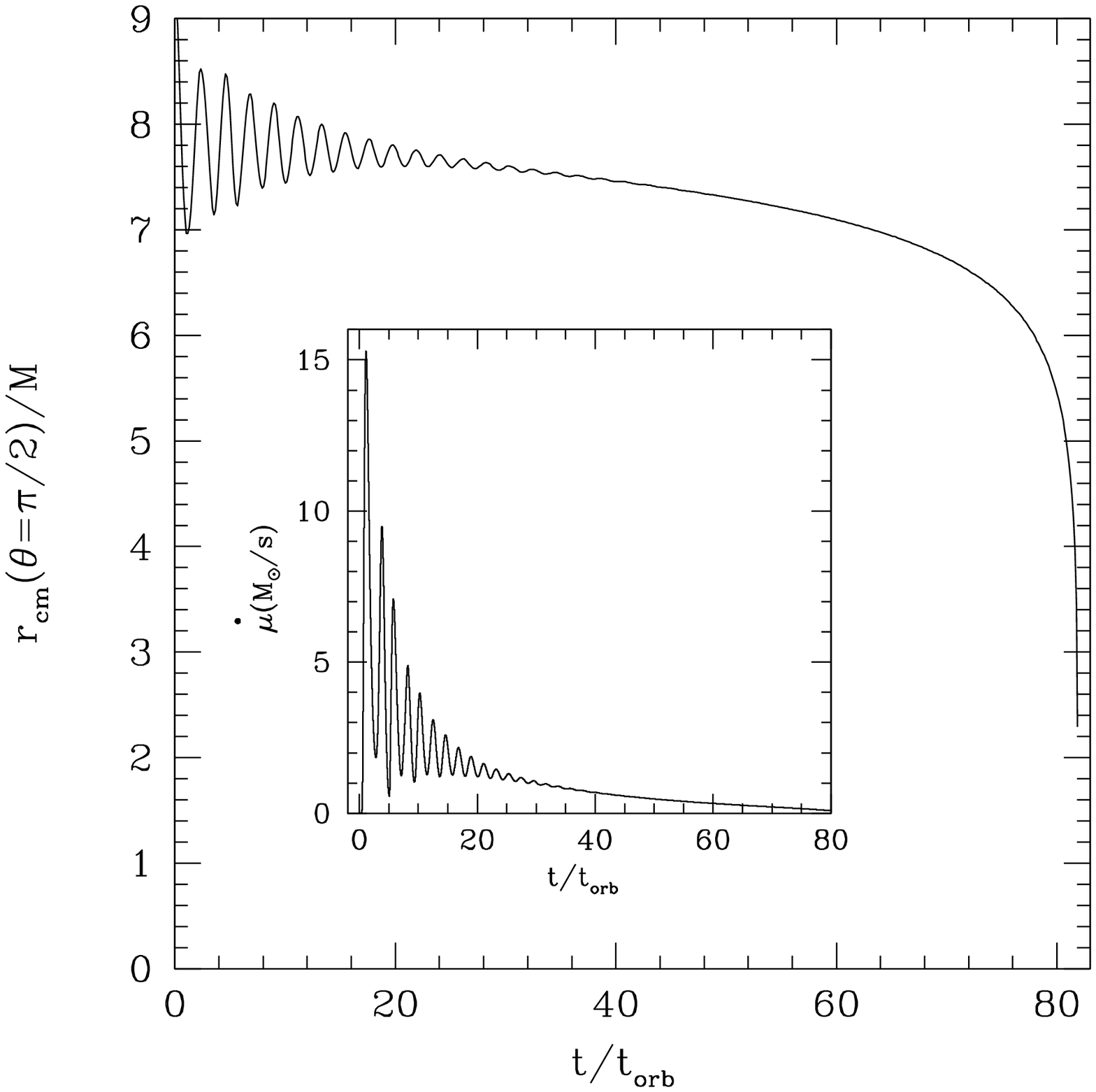}
\hspace{0.5 cm}
\includegraphics[width=8.25 cm]{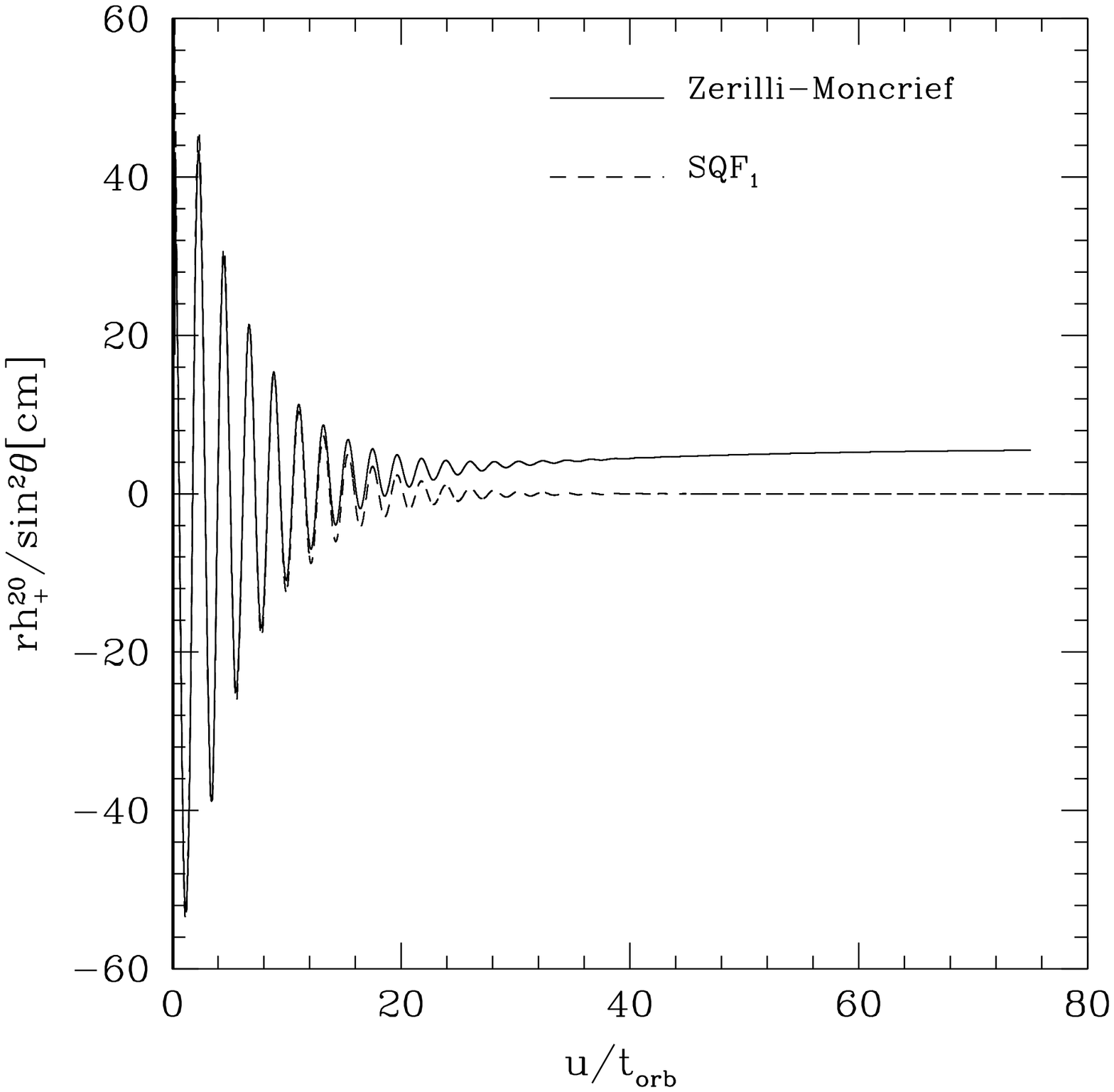}
\vspace{0.0 cm}
\caption{\label{fig13} Model $D_2$. {\it Left panel}: the motion of the
  center of mass on the equatorial plane and the mass accretion rate (in
  the inset). {\it Right panel} : the qudrupole gravitational waveforms.  
  The gauge-invariant (extracted at $r_{\rm obs}=500M$) and the SQF$_1$ 
  waveforms are shown on the same plot for comparison.}
\end{center}
\vspace{-0.5 cm}
\end{figure*}
%%%%%%%%%%%%%%%%%%%%%%%%%%%%%%%%%%%%%%%%%%%%%%%%%%%%%%%%%%%%%%%%%%%%%%%%%

The energy spectrum for this signal is shown as a solid line in the 
right panel of Fig.~\ref{figV_3} and refers to the signal from 
$u/t_{\rm orb}\gtrsim 1$ in order to eliminate the unphysical initial 
burst. The spectrum shows the presence of a broad-band peak at low 
frequencies, related to the highly damped oscillation in the early 
time part of the $\l=2$ signal, together with narrow peaks in the 
harmonic ratio $1:2:3:\dots$, which instead mirror the non-linear coupling
of modes in an oscillating torus (see Refs.~\cite{rezzollaetal_03a,rezzollaetal_03b}
for a detailed discussion of the eigenfrequencies and their
astrophysical implications). Such peaks are present in both the $\l=2$
and $\l=3$ energy spectra (dashed line in Fig.~\ref{figV_3}). The
latter is generated by a  waveform $rh_\times^{30}/(\cos\theta\sin^2\theta)$
whose amplitude is roughly two orders of magnitude smaller than the $\l=2$ one.
It is interesting to note that while for the $\l=2$ multipole the first
frequency of the disk is the dominant one, the second frequency 
has the largest power for the $\l=3$ multipole. The energy spectrum shows
no distinctive signature at $2M\omega \simeq 0.7473$
that could be related with the excitation of the QNMs of the 
black hole (cf., right panel of Fig.~\ref{fig3}). 
This is in contrast with the results of Ref.~\cite{ferrari06}, 
which have simulated a similar source of perturbations for the black hole 
and have indeed found a minute high-frequency contribution in the 
energy spectra related to the emission from the black hole\footnote{It should be 
noted that the QNM contribution is indeed more than four orders of magnitude 
smaller in energy than the one coming from the torus.}. It is presently unclear 
what is the origin of this difference, but it is likely that the hybrid 
approach proposed in Ref.~\cite{ferrari06}, and which combines a solution in
the time-domain for the hydrodynamics and one in the frequency-domain
for the perturbative equations, is better suited to extract the
extremely small contributions coming from the black hole.

A second possibility of triggering quasi-periodic oscillations in the
disk, and hence a quasi-periodic accretion, is by an instantaneous
reduction of the specific angular momentum. The main difference with
the dynamics discussed before is that this reduction produces a
continuous mass accretion, which is however modulated
quasi-periodically as the torus approaches or moves away from the
black hole. This behavior is shown in Fig.~\ref{fig13} for model
$D_2$, in which the specific angular momentum is reduced from
$\l=3.80$ to $\l=3.72$. More specifically, the left panel of
Fig.~\ref{fig13} shows the motion of the center of mass on the
equatorial plane and the evolution of the mass-accretion rate
in the inset. Because of the intrinsic reduction of the centrifugal
support and the resulting continuous accretion, the oscillations are
quasi-periodic in time but with decaying amplitude, so that by
$t/t_{\rm orb} \sim 50$ they have essentially disappeared and the
torus accretes at an almost constant rate, with the center of mass
progressively approaching the black hole.

This different quasi-periodic dynamics of the torus is also imprinted
onto the quadrupolar waveform and this is shown in the right panel of
Fig.~\ref{fig13}, which presents both the gauge-invariant signal as
measured by an observer located at $r_{\rm obs}=500\,M$ (solid line)
and the one computed with the SQF$_{1}$ (dashed line). Note that the
differences between the signals are very small initially, but they
grow with time and by $t\gtrsim 15 t_{\rm orb}$, when the angular
momentum loss induces a global radial motion of the disk, the mean
value of the gauge-invariant signal progressively drifts away from
zero. This different behavior can be understood simply if one bears in
mind that during this stage the disk is undergoing an almost
steady-accretion phase. As a result, while the disks mass-quadrupole
is nonzero and large, its time variations, and hence the corresponding
gravitational-wave signal as computed in the SQF$_1$, are essentially
zero. The perturbative approach, on the other hand, is able to capture
the perturbations induced by this steady-state accretion and this is
reflected in the gradual and secular increase of the
gravitational-wave signal shown in the right panel of
Fig.~\ref{fig13}. Of course, this behavior is the same already
encountered in the case of accretion of dust shells and it is, once
again, generated by curvature backscattering off the ``tail'' of the
potential.

Note that during the final stages of the accretion, i.e., for
$t\gtrsim 60\, t_{\rm orb}$, the motion of the center of mass exhibits
an exponentially rapid motion toward the black hole which, however,
is not reflected neither in the mass-accretion rate, nor in the
gravitational-wave signal. The reason for this is that, in
practice, the rest-mass contained in the torus in these final stages
is extremely small and the rest-mass density is so diluted that it
becomes comparable to that of the atmosphere surrounding the torus. 
As a result, the effective perturbation induced onto the
black hole in the final stages of the accretion is vanishingly 
small.

In summary, the analysis of this process shows that the
gravitational-wave signal produced by a quasi-periodic accretion can
be rather different, depending on whether the amplitude of the
oscillations is (almost) constant in time and the accretion rate is
also periodic, or the oscillations have decreasing amplitude and the
accretion rate is constant with a periodic modulation. In the first
case, the gravitational-wave signal will reflect faithfully the
dynamics of the matter with constant-amplitude waveforms averaging to
zero and a spectrum showing peaks at the eigenfrequencies of the
oscillating matter. In the second case, on the other hand, the
waveforms will have a decreasing amplitude and will not average to
zero as a result of the underlying continuous accretion. 

%-----------
\section{Conclusions}
\label{conclusions}
%-----------

By performing numerical simulations that combine the solution of the
nonlinear hydrodynamics equations with that of the linear
inhomogeneous Zerilli-Moncrief and Regge-Wheeler equations, we have
studied the features of the gravitational-wave signals generated by
the accretion of matter onto a Schwarzschild black hole.  

As extended and accreting matter-sources we have considered
quadrupolar shells of dust falling radially from a finite distance, as
well as geometrically thick disks undergoing either bursts of
hypercritical accretion or quasi-periodic oscillations. In both cases
we find that the gravitational-wave signal \textit{is not} a simple
superposition of the black hole QNMs and that the latter cannot be
found in the energy spectra at times. Rather, we find that quite
generically the signal contains important contributions coming from
radiation scattering off the tail of the curvature potential and
producing a characteristic pattern of interference fringes in the
energy spectra. While the relevance of this contribution differs
according to the specific source considered, it is generically present
as long as the source of perturbations is extended and the scattering
potential does not have an exponential decay with radius. 
This conclusion, which was already reported in previous studies 
involving simpler sources, has been here confirmed unambiguously by 
studying the scattering off a fictitious potential, the P\"oschl-Teller 
potential, which however decays exponentially with radius.

These generic properties of the gravitational-wave emission coming
from black holes perturbed by extended sources represent important
differences with respect to corresponding properties of signals
produced by very compact sources, such as point-like particles. Of
course this conclusion prevents from the derivation of a simple and
generic description of the gravitational-wave signal which would be
independent of the properties of the perturbing source, but it also
opens the exciting perspective of deducing many of the physical
properties of the source through a careful analysis of the waveforms
produced.

Overall, the results presented here make us confident that the
black hole QNM contributions to the full gravitational-wave signal
should be extremely small in generic astrophysical scenarios involving
the accretion of extended distributions of matter. On the other hand,
it should also be stressed that the time-domain analysis carried out
here may not be the most accurate to extract the contributions coming
from the perturbed black hole when these are several orders of
magnitude smaller than those coming from the source itself or from the
backscattering off the potential. In these cases, however, a hybrid
approach such as the one proposed in Ref.~\cite{ferrari06}, combining
the solution in the time-domain for the hydrodynamics with one in the
frequency-domain for the perturbative equations, may be better suited.

%-----------------------------------------------------------------
\begin{acknowledgments}
%-----------------------------------------------------------------

It is a pleasure to thank E.~Berti, H.R.~Beyer, S.~Bernuzzi, 
R.~De Pietri, I.~Olabarrieta, A.~Tartaglia and M.~Tiglio for 
useful discussions and comments. Part of this work was carried out 
by AN through visits in Valencia and at the Center of Computation 
and Technology (CCT) at Louisiana State University; the 
support of the Della Riccia Foundation, of CCT and of ILIAS  is 
gratefully acknowledged. JAF acknowledges financial support from the 
Spanish {\it Ministerio de Educaci\'on y Ciencia} (grant AYA 2004-08067-C03-01).
The computations were performed on the cluster for numerical relativity 
\textit{``Albert''} at the University of Parma.

%-----------------------------------------------------------------
\end{acknowledgments}
%-----------------------------------------------------------------

%%%%%%%%%%%%%%%%%%%%%%%%%%%%%%%%%%%%%%%%%%%%%%%%%%%%%%%%%%

\end{document}